\documentclass[sigconf]{acmart}

\AtBeginDocument{%
  \providecommand\BibTeX{{%
    \normalfont B\kern-0.5em{\scshape i\kern-0.25em b}\kern-0.8em\TeX}}}

\copyrightyear{2021}
\acmYear{2021} 
\setcopyright{iw3c2w3}
\acmConference[WWW '21]{Proceedings of the Web Conference 2021}{April 19--23, 2021}{Ljubljana, Slovenia} 
\acmBooktitle{Proceedings of the Web Conference 2021 (WWW '21), April 19--23, 2021, Ljubljana, Slovenia}
\acmPrice{}
\acmDOI{10.1145/3442381.3450090}
\acmISBN{978-1-4503-8312-7/21/04}

\usepackage{algorithm,algpseudocode,color,graphicx,multirow,paralist,pifont,subfigure,url,xspace,multirow}
\usepackage{balance}

\newcommand*{\scale}[2][4]{\scalebox{#1}{$#2$}}

\setcopyright{iw3c2w3}

\begin{document}

\title{TCN: Table Convolutional Network for Web Table Interpretation}
\fancyhead{}

\author{Daheng Wang$^{1*}$, Prashant Shiralkar$^2$, Colin Lockard$^2$, Binxuan Huang$^2$}
\author{Xin Luna Dong$^2$, Meng Jiang$^1$}
\affiliation{
	\institution{$^1$University of Notre Dame, Notre Dame, IN 46556, USA}
	\institution{$^2$Amazon.com, Seattle, WA 98109, USA}
}
\email{{dwang8, mjiang2}@nd.edu, {shiralp, clockard, binxuan, lunadong}@amazon.com}



\begin{abstract}
Information extraction from semi-structured webpages provides valuable long-tailed facts for augmenting knowledge graph.
Relational Web tables are a critical component containing additional entities and attributes of rich and diverse knowledge.
However, extracting knowledge from relational tables is challenging because of sparse contextual information.
Existing work linearize table cells and heavily rely on modifying deep language models such as BERT which only captures related cells information in the same table.
In this work, we propose a novel relational table representation learning approach considering both the intra- and inter-table contextual information.
On one hand, the proposed Table Convolutional Network model employs the attention mechanism to adaptively focus on the most informative intra-table cells of the same row or column; and, on the other hand, it aggregates inter-table contextual information from various types of implicit connections between cells across different tables.
Specifically, we propose three novel aggregation modules for (i) cells of the same value, (ii) cells of the same schema position, and (iii) cells linked to the same page topic.
We further devise a supervised multi-task training objective for jointly predicting column type and pairwise column relation, as well as a table cell recovery objective for pre-training.
Experiments on real Web table datasets demonstrate our method can outperform competitive baselines by $+4.8\%$ of F1 for column type prediction and by $+4.1\%$ of F1 for pairwise column relation prediction.

\footnotetext[0]{*Most of the work was conducted when the author was interning at Amazon}

\end{abstract}



\settopmatter{printacmref=false, printccs=true, printfolios=false}

\maketitle

{\fontsize{8pt}{8pt} \selectfont \textbf{ACM Reference Format:}\\
Daheng Wang, Prashant Shiralkar, Colin Lockard, Binxuan Huang, Xin Luna Dong, Meng Jiang. 2021. TCN: Table Convolutional Network for Web Table Interpretation. In \textit{Proceedings of the Web Conference 2021 (WWW '21), April 19--23, 2021, Ljubljana, Slovenia.} ACM, New York, NY, USA, 12 pages. https://doi.org/10.1145/3442381.3450090}

\section{Introduction}
\label{sec:introduction}
In recent years, there has been a significant thrust both in academia and industry toward the creation of large knowledge bases (KB) that can power intelligent applications such as question answering, personal assistant, and recommendation. These knowledge bases (e.g., DBpedia \cite{lehmann2015dbpedia}, Wikidata \cite{vrandevcic2014wikidata}, Freebase \cite{bollacker2008freebase}) contain facts about real-world entities such as people, organizations, etc., from a variety of domains and languages in the form of (subject, predicate, object) triples. The field of Information Extraction (IE) aims at populating these KBs by extracting facts from websites on the Web. The information on the Web can be roughly categorized into four types, namely, unstructured text, semi-structured data, Web tables and semantic annotations \cite{multimodalIE2020}. Recently with the advances in natural language processing (NLP), there has been significant progress in the development of effective extraction techniques for the text and semi-structured data \cite{grishman2015information, mausam2016, lockard2019openceres, Lockard2020ZeroShotCeresZR}. However, we have seen limited success to transform the next rich information source, Web tables, into triples that can augment a knowledge base \cite{zhang2019table2vec}. 

\begin{figure}[t]
\centering
{\includegraphics[width=.875\columnwidth]{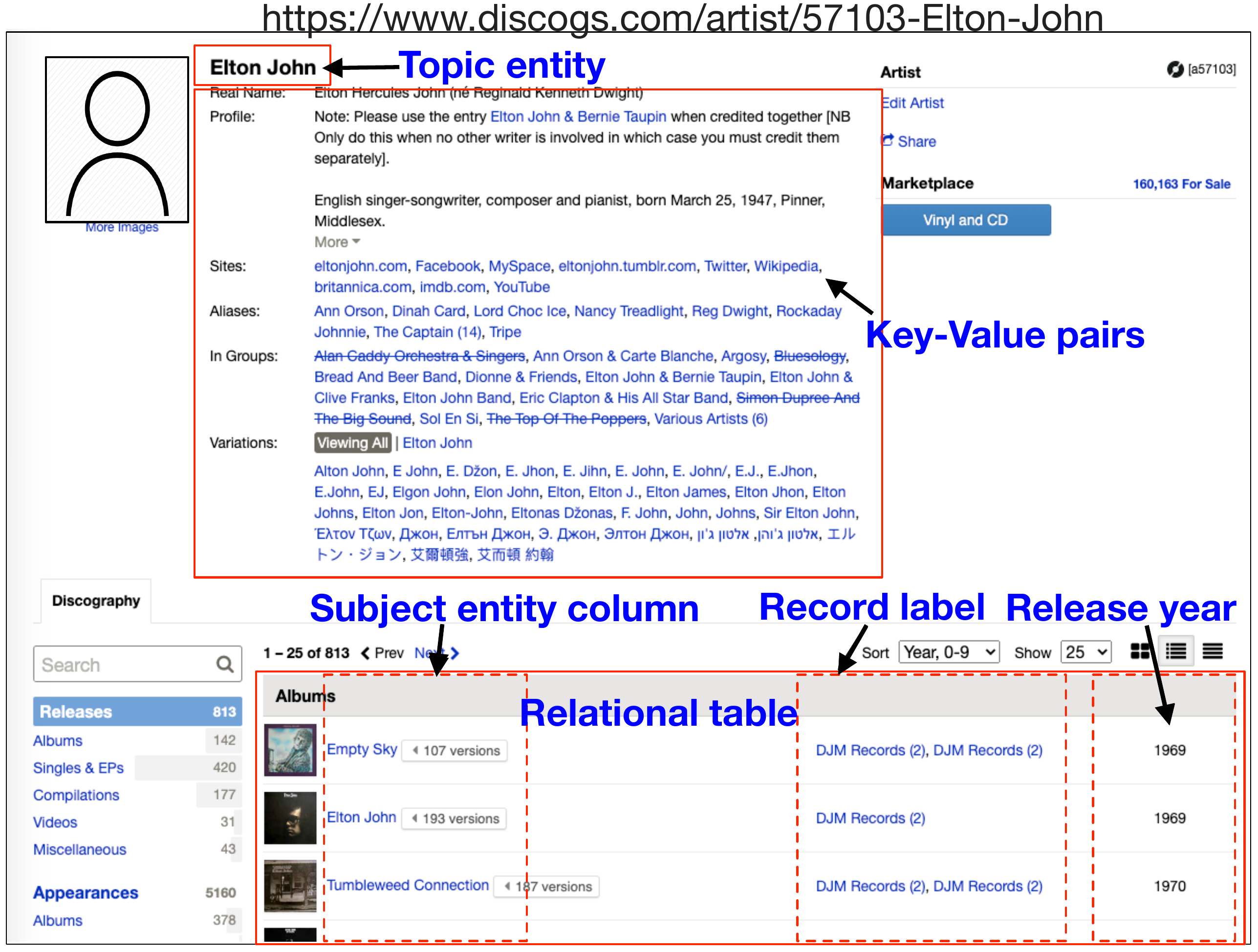}}
    \vspace{-0.1in}
    \caption{An example page of Elton John from \url{discogs.com} showing demographic as key-value pairs and a Web table of discography details. Image is redacted for privacy reasons.}
    \vspace{-0.15in}
    \label{fig:semistructured-elton-john-discogs_v3}
\end{figure}

A Web table is a tabular structure embedded within a webpage displaying information about entities, their attributes and relationships with other entities along rows and columns. It contains high quality relational knowledge and differs from other types of tables such as layout tables, which are primarily meant for formatting purposes, or matrix tables, meant to show numerical summaries in a grid format. Because they contain metadata such as table caption and column headers that mimic tables in a relational database, they are also known as relational Web tables. An example of such table from the detail page of musical artist  ``Elton John'' on \url{discogs.com} is shown in Figure~\ref{fig:semistructured-elton-john-discogs_v3}. This table shows discography information about the artist (the main topic entity of the page) such as albums in which he has performed along the rows, their release date and their publishers, along with other details on the page such as biographical information (e.g., real name, biography, alternate names, etc.) displayed in a key-value format. Although such relational tables are ubiquitous on the Web (a 2016 study estimates a total of 233M tables on the Web \cite{lehmberg2016large}), they are particularly prevalent on semi-structured websites such as \url{discogs.com} which are known to be very rich sources of information in a variety of domains \cite{Lockard2020WebscaleKC}. These sites contain detail pages of different types, e.g., artist, album, track and publisher pages. Since these websites are created by populating HTML templates from an underlying relational database, there are millions of such pages with embedded Web tables, making them particularly attractive for knowledge base enrichment. 

Our goal is to develop effective methods for extracting and interpreting information in Web tables to augment a knowledge base. In this paper, we focus on the task of Web table interpretation, while relying on existing work to perform the task of table extraction, entailing detecting and filtering of tables from the pages \cite{Cafarella2008UncoveringTR}. The task is aligning the schema of a given table to the ontology of a knowledge base. It involves determining the type of each column and the relation between columns from a fixed set of types and relations in the ontology so that tabular entries can be extracted as triples to augment the KB. It is also known as the metadata recovery problem in literature \cite{Cafarella2008UncoveringTR}. However, such alignment is challenging for two reasons, namely \emph{schema heterogeneity} and \emph{context limitation}. The problem of schema heterogeneity arises because tables from different websites use different terminology for table caption and column headers. For example, one website might use ``Name'' while another website might use ``Label'' as the header of a column containing publisher names. Besides, the caption, header and/or tabular cell entry may be missing altogether. The second challenge is that the information in the table cells is often very short, typically consisting of only a few words, and thus lacks adequate context to perform any effective reasoning using off-the-shelf NLP methods, thereby requiring a different approach for table interpretation.

Although the work on Web table interpretation began over a decade ago \cite{cafarella2018ten}, the success has been limited.
Early work employed probabilistic graphical models to capture the joint relationship between rows, columns and table header, but suffered from low precision ($\sim$65\%) \cite{limaye2010annotating}. Most recent work instead ignore the tabular structure and attempt to leverage the power of rich language models such as BERT by applying them to information in the table after transforming it into a long, linear sequence of cell values \cite{yin2020tabert}. However, this approach has the natural downside of resulting in loss of information implicit in the tabular structure, for example, the fact that the values in a column belong to the same type and values along a row belong to the same entity. These implicit connections between cell entries along rows and columns of a table can serve as important context for prediction methods for the two tasks. 

Besides this intra-table context, the tables when viewed as a collection can provide two additional useful \emph{inter-table contexts}. They are (a) shared-schema context: information shared between tables following the same schema from different pages of the same web source. For example, tables from other artist pages of \url{discogs.com} contain some common values such as publisher names producing albums for various artists, and (b) across-schema context: information shared between tables from multiple sources in the same domain, for example, tables from artist pages on both \url{discogs.com} and \url{musicbrainz.org} may show the same publisher names for an album. None of the existing approaches consider the opportunity to leverage this inter-tabular context for designing models for the two tasks.
In this paper, we answer the question: how can we leverage both the intra-table and inter-table contexts to improve web table interpretation, especially in the presence of shared table schema and overlapping values across webpages and even across websites? 

Different from the existing setting, we consider a collection of tables as the input to our problem to utilize the full context available from both intra-table and inter-table implicit connections. We view the collection of tables as a graph in which the nodes represent the tabular cell entries and the edges represent the implicit connections between them. An edge may connect two values from the same row, same column, same cell position in tables following a common schema, or any position in tables having the same value. Given the fact that each node (a value in our case) can have a variable number of links, and inspired by the capabilities of a graph neural network (GNN) to learn effectively from such graph data \cite{wu2020comprehensive, xu2018powerful}, our goal is to learn a table representation that makes use of the intra-table and inter-table contexts for learning a prediction model for the column type and relation prediction tasks. We propose a novel relational table representation learning framework called Table Convolution Network (TCN) that operates on implicitly connected relational Web tables and aggregates information from the available context. 

Our approach gives us two main advantages: (a) it allows us to integrate information from multiple Web tables that provide greater context, and (b) when the inter-table context is unavailable, our model reduces to common case of having only one table as the input, thereby unifying the general case of Web tables. To train the network efficiently, we propose two training schemes: (a) supervised multi-tasking, and (b) unsupervised by way of pre-training for scenarios when the supervision may not be available.

We make the following contributions through this paper:
\begin{compactitem}
	\item We propose a novel representation learning framework called Table Convolution Networks (TCN) for the problem of Web table interpretation involving column type and pairwise column relation prediction. At its core, TCN utilizes the intra-table and inter-table context available from a collection of tables, instead of the limited context from a single table.
	\item We show two approaches to train the network, namely a classic supervised mode, and an unsupervised mode that employs pre-training through self-supervision for jointly predicting column type and relation between column pairs.
	\item We perform extensive experiments with several state-of-the-art baselines on two datasets containing 128K tables of 2.5M triples, showing that TCN outperforms all of them with an F1 score of 93.8\%, a relative improvement of 4.8\% points on the column type detection task, and an F1 score of 93.3\%, a relative improvement of 4.1\% on the relation prediction task.
\end{compactitem}

The roadmap of this paper is organized as follows. We review related work in Section~\ref{sec:related}. In Section~\ref{sec:problem}, we formally define the research problem. Our proposed Table Convolutional Network approach is introduced in Section~\ref{sec:approach}. Section~\ref{sec:experiments} presents experimental results. We conclude the paper and discuss on future work in Section~\ref{sec:conclusions}.

\section{Related Work}
\label{sec:related}
We discuss two lines of research related to our work.

\vspace{0.05in}
\noindent \textbf{Relational Table Interpretation.}
Relational tables on the Web describe a set of entities with their attributes and have been widely used as vehicle for conveying complex relational information \cite{cafarella2008webtables, fetahu2019tablenet}.
Since the relationships of table cells are not explicitly expressed, relational table interpretation aims at discovering the semantics of the data contained in relational tables, with the goal of transforming them into knowledge intelligently processable by machines \cite{cafarella2018ten, yu2019tablepedia}.
With the help from existing knowledge bases, this is accomplished by first classifying tables according to some taxonomy \cite{nishida2017understanding,ghasemi2018tabvec}, then identifying what table columns are about and uncovering the binary relation of table columns \cite{zhang2020web}.
The extracted knowledge can in turn be readily used for augmenting knowledge bases \cite{ritze2016profiling}.

Column type annotation refers to associating a relational table column with the type of entities it contains.
Earlier methods combine the exact match or certain entity search strategies with a majority vote scheme for predicting the column type \cite{mulwad2010using, venetis2011recovering}.
Fan et al. \cite{fan2014hybrid} proposed a two-stage method which first matches column to candidate concepts and then employ crowdsourcing for refinement on type prediction.
\textsc{T2KMatch} by Lehmberg et al. \cite{lehmberg2017stitching} proposed to stitch Web tables of the same schema into a larger one to improve the prediction performance.
\textsc{Sherlock} \cite{hulsebos2019sherlock} by Hulsebos et al. proposed a set of statistical features describing the character and word level distributions along with some semantic features for feeding into a deep classifier to get high prediction accuracy.

Relation extraction is the task of associating a pair of columns in a table with the relation that holds between their contents.
Mulwad et al. \cite{mulwad2013semantic} proposed a semantic message passing algorithm using knowledge from the linked open data cloud to infer the semantics between table columns.
Munoz et al. \cite{munoz2014using} proposed to use an existing linked data knowledge base to find known pre-existing relations between entities and extend on analogous table columns.
Sekhavat et al. \cite{sekhavat2014knowledge} proposed a probabilistic model leveraging natural language patterns associated with relations in knowledge base.

Another common task for interpreting relational table is entity linking which is the process of detecting and disambiguating specific entities mentioned in the table \cite{efthymiou2017matching, bhagavatula2015tabel}.
Existing work often couple it with column type annotation and relation extraction together as a prerequisite or joint task \cite{mulwad2010using, ritze2015matching, zhang2017effective}.
However, this requires expensive preprocessing steps and largely limits the flexibility for interpreting semantics of relational table \cite{chen2019learning}.
In this work, we do not assume the availability of any pre-linked entities, and focus on the tasks of column type annotation and pairwise column relation extraction completely based on the table cell contents.

\vspace{0.05in}
\noindent \textbf{Representation Learning of Tabular Data.}
Earlier work utilized probabilistic models to capture dependencies between table cells.
Limaye et al. \cite{limaye2010annotating} proposed to model the entity, type and relation information of table cells as random variables, and jointly learn their distributions by defining a set of potential functions.
A Markov Random Fields model was proposed by Ibrahim et al. \cite{ibrahim2016making} for canonicalizing table headers and cells into concepts and entities with a special consideration on numerical cell values of quantities.
\textsc{Meimei} by Takeoka et al. \cite{takeoka2019meimei} proposed to incorporate multi-label classifiers in the probabilistic model to support versatile types of cell and improve predictive preformance.
These methods have high complexity due to MCMC sampling and they cannot be directly applied on large-scale relational Web tables.

Some studies made efforts in table representations learning by leveraging the word embedding model \textsc{word2vec} \cite{mikolov2013distributed}.
\textsc{Table2Vec} by Zhang et al. \cite{zhang2019table2vec} proposed to linearize a cropped portion of the table's grid structure into sequence of cell tokens as the input of \textsc{word2vec}.
This treatment was also adopted by Gentile et al. \cite{gentile2017entity} for blocking to reduce the efficiency of entity matching.
\textsc{Record2Vec} by Sim et al. \cite{sim2018record2vec} transformed structured records into attribute sequence and combined \textsc{word2vec} with a tailored triplet loss.
However, shallow neural models like \textsc{word2vec} have relatively limited expressiveness which pose difficulties on effectively capturing the semantics of relational tables.

Recent methods utilize deep neural language model for learning table representations.
\textsc{TURL} by Deng et al. \cite{deng2020turl} proposed a pre-training/finetuning framework for relational Web tables by injecting visibility matrix into the encoder of \text{Transformer} \cite{vaswani2017attention} to attend on structurally related table components. The authors also proposed a Masked Entity Recovery objective to enhance the learning capability but this requires pre-linked entities of table cells as model input which is not available in most real cases.
\textsc{TaPas} by Herzig et al. \cite{herzig2020tapas} proposed to jointly learn the embedding of natural language questions over relational tables by extending the \textsc{BERT} \cite{devlin2018bert} model with more table-aware positional embeddings.
\textsc{TaBERT} by Yin et al. \cite{yin2020tabert} adopted a similar idea for semantic parsing on database tables combining with Masked Column Prediction and Cell Value Recovery as two additional unsupervised objectives for pre-training.
One major limitation of these methods is they only focus on aggregating components of a single table via indirect techniques such as visibility matrix and content snapshot \cite{nguyen2019mtab, cremaschi2019mantistable, steenwinckel2019csv2kg}.
In contrast, we propose to directly capture intra-table context by attending on the column and rows cells with easy integration of inter-table contexts. And, we fully consider the highly valuable inter-table contextual information by aggregating various types of implicit connections between tables of same cell value or position.

\section{Problem Definition}
\label{sec:problem}
\begin{table}[t]
	\renewcommand{\arraystretch}{1.15}
	\centering
	\caption{Symbols and their descriptions.}
	\label{tab:notations}
	\vspace{-0.1in}
	\scale[0.925]{
	\begin{tabular}{|c||l|}
	\hline
		\textbf{Symbol} & \textbf{Description} \\ \hline \hline
		$T_k$ & a relational Web table \\ \hline
		\multirow{2}*{$t^{m,n}$ ($t^{m,n}_k$)} & table cell of $T_k$ locates at the intersection of \\
		 & the $m$-th row and the $n$-th column \\ \hline
		$t^{m,*}$, $t^{*,n}$ & the $m$-th row, and the $n$-th column of $T_k$ \\ \hline
		$S^r_k$, $S^c_k$ & $T_k$'s number of rows, and number of columns \\ \hline
		$\phi$ & the table schema mapping function \\ \hline
		$p_k$ & $T_k$'s page topic of short text \\ \hline
		$\mathcal{D}$ & a dataset of relational tables \\ \hline
		\multirow{2}*{$K$, $U$} & number of relational tables, and number of \\
		 & unique table schema in $\mathcal{D}$ \\ \hline
		\multirow{2}*{$\mathcal{C}$, $\mathcal{R}$} & set of target column types, and set of target  \\
		 & relations between subject and object columns \\ \hline
		$\mathbf{e}_{t^{m,n}}$ & initial embedding vector of table cell $t^{m,n}$ \\ \hline
		\multirow{2}*{$\mathbf{e}_{t^{m,n}}^c$, $\mathbf{e}_{t^{m,n}}^r$} & the column-wise, and row-wise aggregated \\
		 & context vectors of target cell $t^{m,n}$ \\ \hline
		$\text{AGG}_a$ & the intra-table aggregation function \\ \hline
		$\mathbf{e}_{t^{m,n}}^a$ & the intra-table contextual embedding of $t^{m,n}$ \\ \hline
		\multirow{2}*{$\mathcal{N}_{v}$, $\mathcal{N}_{s}$, $\mathcal{N}_{p}$} & set of value cells, set of position cells, and \\
		 & set of and topic cells \\ \hline
		$\mathbf{e}^{v}_{t^{m,n}}$, $\mathbf{e}^{s}_{t^{m,n}}$ & aggregated inter-table contextual embeddings \\
		$\mathbf{e}^{p}_{t^{m,n}}$ & of $\mathcal{N}_{v}(t^{m,n})$, $\mathcal{N}_{s}(t^{m,n})$, and $\mathcal{N}_{p}(t^{m,n})$ \\ \hline
		$D$, $\mathbf{W}$ & dimension of vector, and matrix of parameters \\ \hline
	\end{tabular}
	}
	\vspace{-0.2in}
\end{table}

Given a collection of webpages, perhaps from semi-structured websites, with relational Web tables embedded in them, and our goal is to uncover the semantics of the columns by annotating them with their type and determining the relation between pairs of columns. 
A Web table in such pages can be schematically understood as a grid comprising of rows and columns. Each row contains information about a single real-world entity (e.g., an album name), typically found in one of the first few columns, and its attributes and relationships with other entities in other columns (e.g., release year and publisher), and each column contains attributes or entities of the same type described by an optional column header.
Taking the table in Figure~\ref{fig:semistructured-elton-john-discogs_v3} as an example, the first column contains the album entities being described, while the rest of the columns indicate its attributes and relationships. We call this column the \emph{subject} column to indicate it is the subject of the rows. Moreover, we can infer ``DJM Records'' to be the publisher of ``Empty Sky'' due to the fact that their association is found in the same row, and likewise, we can inter ``Empty Sky'' to be a ``Release'' (a technical term for album) by knowing other values in the same column and due to presence of the header ``Album''. In a set of $K$ relational Web tables, we denote the $k$-th table $T_k$ $(k=1,\dots,K)$ as a set of row tuples, i.e., $T_k \coloneqq \{(t^{0,0}_k,t^{0,1}_k,\dots,t^{0,S^c_k}_k), \dots ,\allowbreak (t^{S^r_k,0}_k,t^{S^r_k,1}_k,\dots,t^{S^r_k,S^c_k}_k)\}$ where $S^r_k$ and $S^c_k$ are the number of rows and columns of the $k$-th table. The first row $t^{0,*}$ typically contains the table header (e.g., ``Title of Track'' and ``Name of Composer''). When the context is clear, we omit the subscript and use $t^{m,n}$ to denote the cell at the intersection of $m$-th row ($0 \leq m \leq S^r_k$) and $n$-th column ($0 \leq n \leq S^c_k$) of table $T_k$. We use $t^{m,*}$ and $t^{*,n}$ to denote all cells at the $m$-th row and the $n$-th column of the table respectively, i.e., $t^{m,*} \coloneqq (t^{m,0},t^{m,1},\dots,t^{m,S^c_k})$ and $t^{*,n} \coloneqq (t^{0,n},t^{1,n},\dots,t^{S^r_k,n})$.

A Web table has additional contexts that further describe its semantics and therefore should be leveraged for better table interpretation. They come in two forms: metadata from the page of the table, namely an optional table caption and the \texttt{<title>} tag or topic entity of the webpage, and an inter-table context available by considering the collection of tables that either conform to the same underlying schema by virtue of belonging to the same HTML template, or describe similar information from different sites in the same domain. Being present on the detail page of "Elton John" (the topic entity in our example), the table conveys the semantic connection between him and the set of albums in the table, thereby providing additional context for the cell values. 

Furthermore, there also exists greater context by considering the implicit connections between tables as a collection. We assume pages of a website have been segregated to form groups that correspond to distinct HTML templates conforming to different page types (i.e., schema) thus each table belongs to a single unknown schema. Accordingly, cell entries in the same position across tables of the same schema provide evidence of belonging to the same type. This can be particularly useful when one of the tables is sparse and difficult to reason about. We can also view the connections between same values from different tables, perhaps from entirely disparate schema, to also provide additional useful information, e.g., one table from site \textit{A} might describe a publisher's information with artists, while another table from site \textit{B} might describe its information with bands. Such inter-tabular context can be valuable to discern subtle differences for the task of column type and relation prediction.


We use $\phi:\{1,\dots,K\} \rightarrow \{1,\dots,U\}$ to denote a table schema mapping function where $U$ is the number of unique schema. Relational tables of the same schema $\{T_k \,|\, \phi(k)=u, 1 \leq u \leq U \}$ have the same header cells and generally $U$ is a large value comparable to $K$. In practice, when relational tables are from semi-structured websites focusing around certain page topics, the value of $U$ could be much smaller than $K$, i.e., $U << K$.
Besides, table cells of the same column can be characterized by the the common known entity type (e.g.,  ``Release'' for tracks and ``People'' for composers). Assuming the first column $t^{*,0}$ contains all subject entities (e.g., various tracks of an album) we denote it as the \emph{subject} column, and all other columns of the table $t^{*,n}$ ($1 \leq n \leq S^c_k$) contain attribute information about the first column, we denote them as the \emph{object} columns. Each relational table $T_k$ describes a set of pairwise relations between different pairs of subject column and object column $(t^{*,0}, t^{*,n})$ where $n=1,\dots,S^c_k$. In addition, each relational table $T_k$ is also accompanied with a page topic $p_k$. This is usually a short phrase (e.g., the page title) summarizing the table's main theme (e.g., the name of the artist performing in the music albums).

We now formally define our research problem as below:

\vspace{0.03in} 
\noindent \textbf{Problem:} Given a relational table dataset $\mathcal{D}=\{(T_k,p_k)\}^{K}_{k=1}$, our goal is to perform the following two table interpretation tasks:

\vspace{0.03in} 
\noindent \textbf{Column type detection:} we aim to predict the type of a column from among a fixed set of predefined types from an ontology, i.e., $f_c: \{\{t^{*,n}_k\}^{S^c_k}_{n=0}\}^{K}_{k=1} \rightarrow \mathcal{C}$ where $\mathcal{C}$ is the set of target entity types;  


\vspace{0.03in} 
\noindent \textbf{Pairwise column relation prediction:} we aim to predict the pairwise relation between the subject column and object columns, i.e., $f_r: \{\{(t^{*,0}_k,t^{*,n}_k)\}^{S^c_k}_{n=1}\}^{K}_{k=1} \rightarrow \mathcal{R}$ where $\mathcal{R}$ is the set of known relations from the known ontology.


\section{The Proposed Approach}
\label{sec:approach}
\begin{figure*}[t]
	\centering
	{\includegraphics[width=0.85\textwidth]{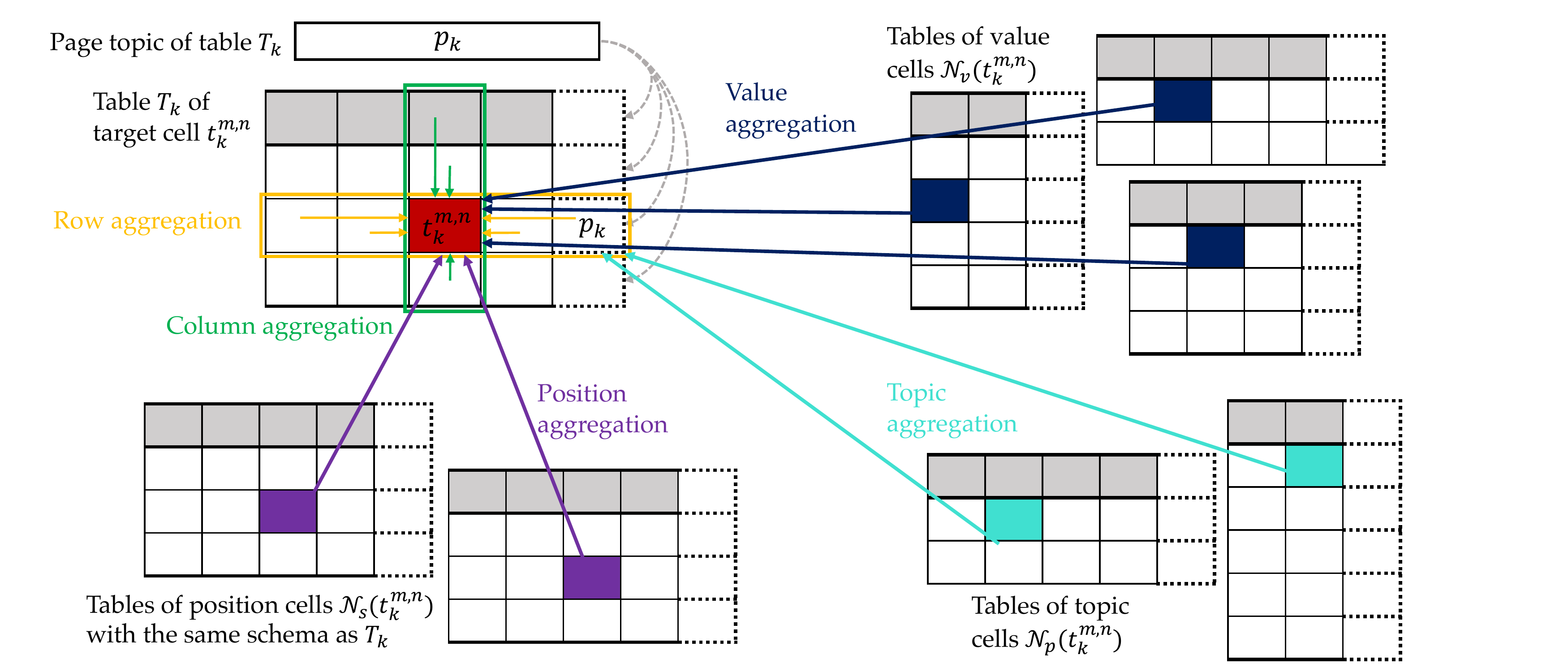}}
	\vspace{-0.1in}
	\caption{Overall framework of the proposed \textsc{TCN} for learning relational table latent representations by considering both the intra- and inter-table contextual information. Page topic $p_k$ is appended to the right of each table as a pseudo-column (dashed cells). Arrows/cells highlighted in various colors denote different types of connection to the target cell $t^{m,n}_k$ and the corresponding aggregation module. The intra-table context of $t^{m,n}_k$ is aggregated from cells of the same column and row (green and yellow). Morevoer, 3 types of inter-table contexts are aggregated from (i) value cells $\mathcal{N}_{b}$ of the same value (blue), (ii) position cells $\mathcal{N}_{s}$ of the same schema position (purple), and (iii) topic cells $\mathcal{N}_{p}$ of the same value as $t^{m,n}_k$'s topic (turquoise) respectively.}
	\vspace{-0.15in}
	\label{fig:framework}
\end{figure*}

In this section, we present a novel deep architecture Table Convolutional Network (\textsc{TCN}) for relational table representation learning.
\textsc{TCN} first learns the latent embedding of each relational table cell by aggregating both intra- and inter-table contextual information. These learned cell embeddings are then summarized into column embedding which are used for predicting the column type and pairwise column relation.
The overall framework of \textsc{TCN} is shown in Figure \ref{fig:framework}.
We first introduce the intra-table aggregation module for summarizing cells of the same column and row (Section \ref{subsec:approach_agg_intra}).
Then, for capturing the contextual information across tables, we propose three specific inter-table aggregation methods to fully learn from various types of implicit connections between tables (Section \ref{subsec:approach_agg_inter}).
At last, we present the model's training procedure in a classic supervised setting as well as for pre-training on large-scale relational Web tables dataset (Section \ref{subsec:approach_obj}).

\subsection{Intra-table Aggregations}
\label{subsec:approach_agg_intra}
For learning the latent representation of a target table cell $t^{m,n} \in T_k$, besides the information carried by its own cell value, it is natural to assume other cells in $T_k$ of the same column or row are helpful for capturing the intra-table context of $t^{m,n}$.
As an example, a single cell that of a common person name ``Pete Bellotte'' is ambiguous by itself, unless other song composer names appear in the same column, or his composed song names are present in the same row.

\subsubsection{Column aggregation}
We use $\mathbf{e}_{t^{m,n}} \in \mathbb{R}^{D_d}$ to denote the initial $D_d$-dim embedding vector of cell $t^{m,n}$. It can be pre-trained word embeddings \cite{pennington2014glove} or simply setting to one-hot identifier vector.
A straightforward way to consider other cells of the same column as context of $t^{m,n}$ is applying a pooling operator on their embeddings, e.g., $\sum^{S^r_k}_{m'=0} \mathbf{e}_{t^{m',n}} / (S^r_k-1)$ where $m' \neq m$.
However, different cells in the column have various contributions to the context of the target cell and they should be considered differently.
For example, in a trending songs table of a singer, cells of his or her main artist songs should be more important of larger weight values compared with featured artist songs. 
This can be achieved by setting the target cell embedding $\mathbf{e}_{t^{m,n}}$ as query to attend on other cell embeddings $\{\mathbf{e}_{t^{m',n}}\}^{S^r_k}_{m'=0}$ $(m'\neq m)$ of the same column (see Figure~\ref{fig:agg_column}):
\begin{equation}
	\label{eqn:agg_intra_column_weights}
	\alpha_{t^{m',n}} = \frac{\text{exp} \left( {\mathbf{e}_{t^{m',n}}}^\top \cdot \mathbf{e}_{t^{m,n}} \right) }{\sum^{S^r_k}_{\tilde{m}=0, \tilde{m}\neq m} \text{exp}\left( {\mathbf{e}_{t^{\tilde{m},n}}}^\top \cdot \mathbf{e}_{t^{m,n}} \right)},
\end{equation}
where $\alpha_{t^{m',n}}$ is the weight of column cell $t^{m',n}$. The column-wise aggregated context embedding $\mathbf{e}^{c}_{t^{m,n}} \in \mathbb{R}^{D_c}$ can be computed by
\begin{equation}
	\label{eqn:agg_intra_column}
	\mathbf{e}^{c}_{t^{m,n}} = \sigma \left( \mathbf{W}_c \cdot \sum^{S^r_k}_{m'=0,m'\neq m} \alpha_{t^{m',n}} \mathbf{e}_{t^{m',n}} \right),
\end{equation}
where $\sigma$ is nonlinear {ReLU}, $\mathbf{W}_c \in \mathbb{R}^{D_c \times D_d}$ is parameter matrix.

\subsubsection{Row aggregation}
\label{subsubsec:approach_agg_intra_row}
\begin{figure}[t]
    \centering
    \subfigure[Column aggregation]{
    	\includegraphics[width=0.475\linewidth]{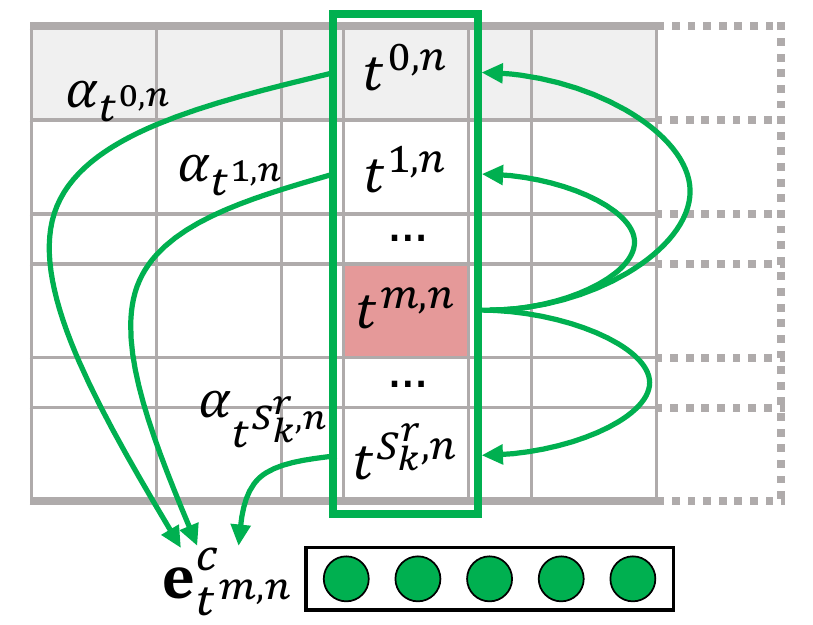}
		\label{fig:agg_column}}
    \hfill
    \subfigure[Row aggregation]{
    	\includegraphics[width=0.475\linewidth]{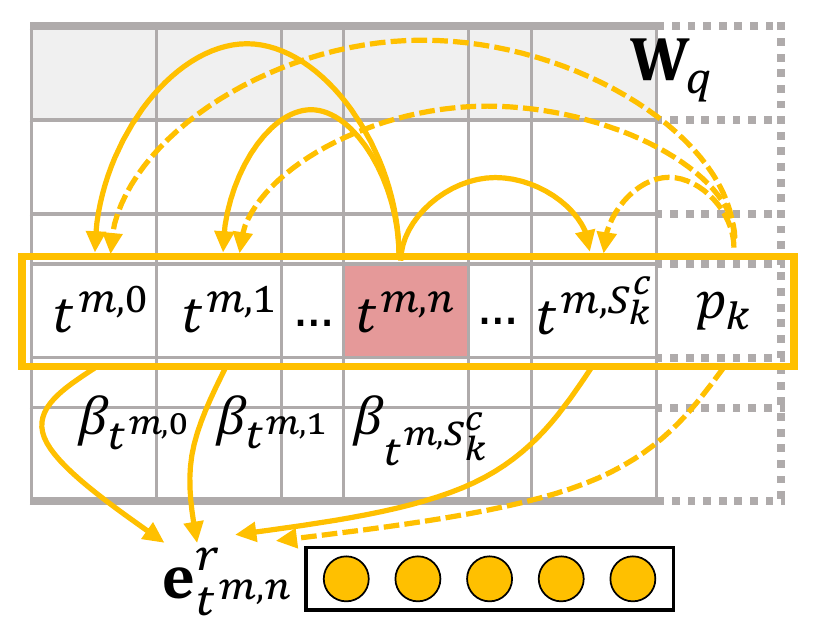}
		\label{fig:agg_row}}
	\vspace{-0.1in}
    \caption{The intra-table aggregation module for summarizing contexts of target cell $t^{m,n}$ inside table $T_k$. (a) Column aggregation uses the embedding of $t^{m,n}$ as query to attend on other column cells for generating column-wise aggregated context $\mathbf{e}_{t^{m,n}}^c$. (b) Row aggregation also incorporates the embedding of page topic $p_k$ into the query for attending other row cells. The result is concatenated with topic embedding as the row-wise aggregated context embedding $\mathbf{e}_{t^{m,n}}^r$.}
    \label{fig:agg_intra}
    \vspace{-0.1in}
\end{figure}

Analogous to column aggregation, we can also use the target cell as query to attend and aggregate other row cells.
However, different from column cells that are homogeneous of the same entity type, cells of the same row are mostly heterogeneous in type and contain complex relational information conditioning on the page topic $p_k$ \cite{zhang2017effective}. In other words, knowing the page topic can greatly benefit inferring the factual knowledge of other row cells with respect to $t^{m,n}$.
For capturing the impact from page topic $p_k$, we incorporate the topic embedding vector $\mathbf{e}_{p_k} \in \mathbb{R}^{D_p}$ into the target cell query $\mathbf{e}_{t^{m,n}}$ for attending other row cells (see Figure~\ref{fig:agg_row}):
\begin{equation}
	\label{eqn:agg_intra_row_weights}
	\beta_{t^{m,n'}} = \frac{\text{exp} \left( {\mathbf{e}_{t^{m,n'}}}^\top \cdot \mathbf{W}_q \cdot (\mathbf{e}_{t^{m,n}} \| \mathbf{e}_{p_k}) \right) }{\sum^{S^c_k}_{\tilde{n}=0, \tilde{n}\neq n} \text{exp}\left( {\mathbf{e}_{t^{m,\tilde{n}}}}^\top \cdot \mathbf{W}_q \cdot (\mathbf{e}_{t^{m,n}} \| \mathbf{e}_{p_k}) \right)},
\end{equation}
where $\mathbf{W}_q \in \mathbb{R}^{D_d \times (D_d+D_p)}$ is a bilinear transformation allowing interactions from row cells to both the target cell and page topic, and $\|$ is the vector concatenation operator. So, comparing with Eqn.~(\ref{eqn:agg_intra_column_weights}) the attention weights of row cells are adaptively determined based on the target cell information as well as the page topic semantics.

In addition, we explicitly include the page topic into the row-wise aggregated context vector by concatenating the topic embedding $\mathbf{e}_{p_k}$ with the attended sum of row cell embeddings:
\begin{equation}
	\label{eqn:agg_intra_row}
	\mathbf{e}^{r}_{t^{m,n}} = \sigma \left( \mathbf{W}_r \cdot \sum^{S^c_k}_{n'=0,n'\neq n} \left( \beta_{t^{m,n'}} \mathbf{e}_{t^{m,n'}} \right) \| \mathbf{e}_{p_k} \right),
\end{equation}
where $\mathbf{e}^{r}_{t^{m,n}} \in \mathbb{R}^{D_r}$ denotes the row-wise aggregated $D_r$-dim context embedding of $t^{m,n}$ and $\mathbf{W}_r \in \mathbb{R}^{D_r \times (D_d+D_p)}$ is parameter matrix.
Intuitively, this can be seen as appending the page topic $p_k$ as a pseudo-column of identical topic cells to the last column of $T_k$.

\subsubsection{The intra-table aggregation module}
\label{subsubsec:approach_agg_intra_func}
After we have distilled contextual information of $t^{m,n}$ by aggregating from related cells of the same column and row in $T_k$, we can fuse these column- and row-wise aggregated context embeddings into a holistic intra-table context embedding $\mathbf{e}^{a}_{t^{m,n}} \in \mathbb{R}^{D_a}$.
We use function $\text{AGG}_a$ to denote this whole intra-table aggregation process (from Eqn.~(\ref{eqn:agg_intra_column}) to (\ref{eqn:fuse_intra})):
\begin{equation}
	\label{eqn:fuse_intra}
	\mathbf{e}^{a}_{t^{m,n}} = \sigma \left( \mathbf{W}_a \cdot (\mathbf{e}^{c}_{t^{m,n}} \| \mathbf{e}^{r}_{t^{m,n}}) \right) = \text{AGG}_a(t^{m,n}),
\end{equation}
where $\mathbf{W}_a \in \mathbb{R}^{D_a \times (D_c+D_r)}$ is the parameter matrix.
The output embedding $\mathbf{e}^{a}_{t^{m,n}}$ encapsulates the intra-table contextual information of target cell $t^{m,n}$ from all informative cells of relational table $T_k$.
Most existing work of table representation learning rely on indirect techniques such as visibility matrix \cite{deng2020turl} and content snapshot \cite{yin2020tabert} for modeling related cells inside the table and does not consider contexts across tables.
In contrast, the proposed intra-table aggregation of \textsc{TCN} directly captures the intra-table context, and can be easily applied for integrating with various inter-table contexts.
We use this intra-table aggregation function $\text{AGG}_a$ as the underlying operation for summarizing all intra-table contexts of arbitrary cells that are implicitly connected to the target cell $t^{m,n}_k$.

\begin{figure}[t]
    \centering
    {\includegraphics[width=1\columnwidth]{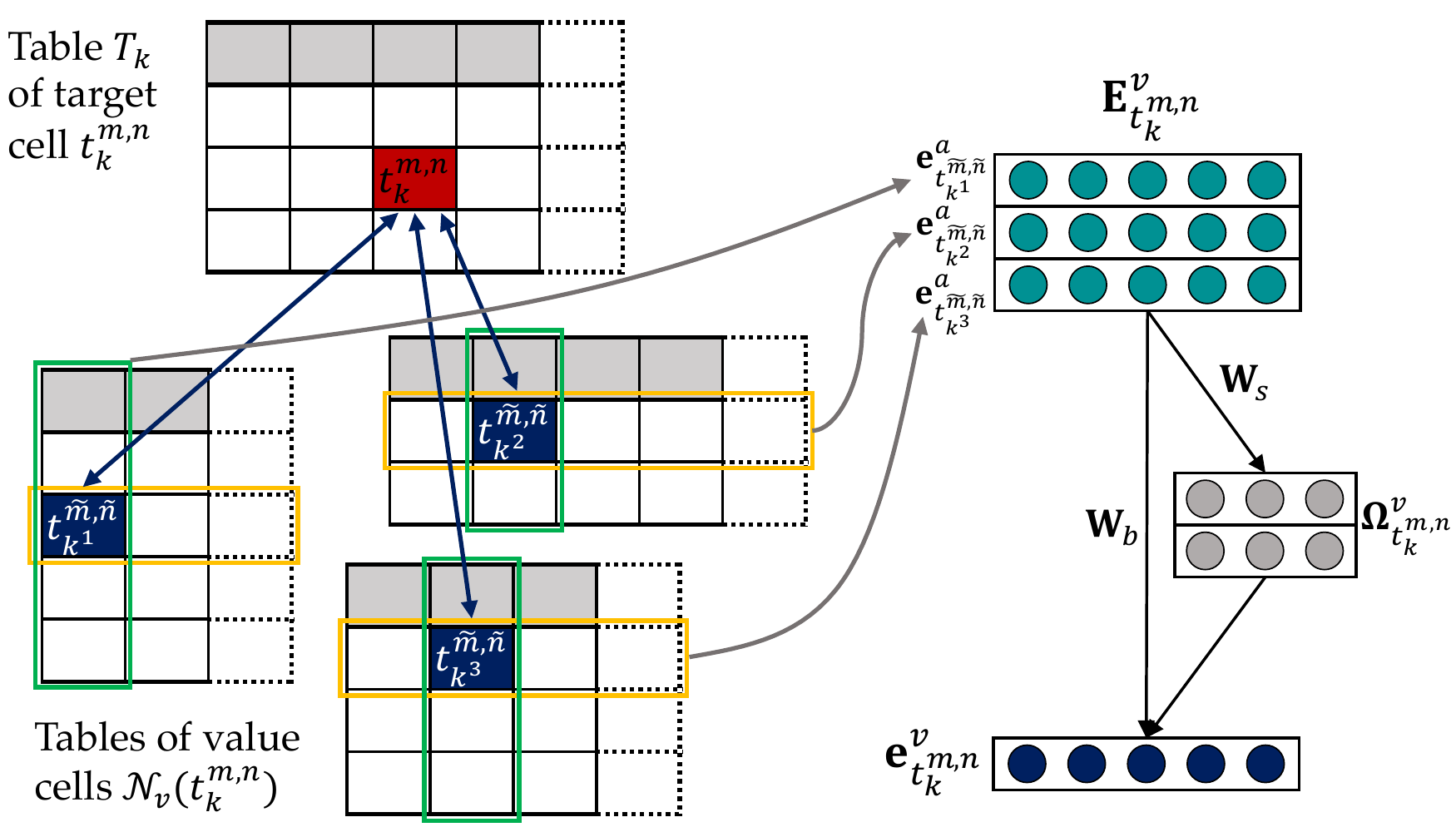}}
    \vspace{-0.15in}
    \caption{The value aggregation module for summarizing the target cell $t^{m,n}_k$'s inter-table contextual information from its value cells $\mathcal{N}_{v}(t^{m,n}_k)$. Double-sided arrows indicate $\mathcal{N}_{v}$ share the same cell value as $t^{m,n}_k$. The intra-table contexts of $\mathcal{N}_{v}$ extracted via $\text{AGG}_a$ (Section~\ref{subsubsec:approach_agg_intra_func}) are arranged into matrix $\mathbf{E}^{v}_{t^{m,n}}$. The value cells aggregated context embedding $\mathbf{E}^{v}_{t^{m,n}}$ is summed by self-attention weights of $\mathbf{\Omega}^{v}_{t^{m,n}}$.}
    \vspace{-0.15in}
    \label{fig:agg_inter_neighbor}
\end{figure}

\subsection{Inter-table Aggregations}
\label{subsec:approach_agg_inter}
By aggregating from related cells in the same table, we can learn locally context-aware latent representation of a table cell.
However, on the Web there are also a lot of \emph{implicit} connections across different tables, and these hidden connections often provide highly valuable context that are complementary to the intra-table context.
For example, the song composer's name ``Pete Bellotte'' could appear in multiple tables where he also serves as a record producer in some of them. These two roles are subtly different yet complementary to each other.
Jointly modeling the intra- and inter-table contexts can benefit capturing more accurate relational information.

Such inter-table connections can also be of various types. Besides tables connected by the same cell value, there are often tables sharing the same schema (i.e., the same headers), and the topic of certain tables might appear in other tables as cell values.
For example, Web tables designed for describing music albums will have the identical header cells, and the page topic (i.e., the album name) can also appear in the singer's discography table.
To effectively modeling context of heterogeneous connections, we propose three inter-table aggregation modules for distilling inter-table contexts.

\subsubsection{Value aggregation}
\label{subsubsec:approach_agg_inter_neighbor}
Each relational table describe a set of relations between its cells, and in turn each unique cell could also be expressed by a set of tables where it appears. Intuitively, each table can be seen as a partial view of target cell's context.
For capturing the contextual information from different tables, we establish connections between cells of different tables containing the same value. Particularly, we adopt basic normalizing procedures to canonicalize table cells with no additional step of expensive entity linking \cite{mulwad2010using, ritze2015matching, zhang2017effective}. In practice, we apply minimal preprocessing on cell values by lowering all cases and removing redundant spaces. 
Given a target cell $t^{m,n}_k$ in relational table $T_k$, we use $\mathcal{N}_{v}$ to denote cells of other tables containing the same value, i.e., $\mathcal{N}_{v}(t^{m,n}_k) \coloneqq \{t^{\tilde{m},\tilde{n}}_{k'}~|~t^{\tilde{m},\tilde{n}}_{k'}=t^{m,n}_k \wedge 0 \leq k' \leq K \wedge k'\neq k\}$.

By applying the intra-table aggregation function $\text{AGG}_a$ as previously introduced (Section~\ref{subsubsec:approach_agg_intra_func}), we can produce the local contexts of all value cells $\{ \mathbf{e}^{a}_{t^{\tilde{m},\tilde{n}}_{k'}}=\text{AGG}_a(t^{\tilde{m},\tilde{n}}_{k'}) \,|\, t^{\tilde{m},\tilde{n}}_{k'} \in \mathcal{N}_{v}(t^{m,n}_k) \}$ with respect to the corresponding relational table.
For effectively focusing on the most useful connections of value cells, we further process this variant-sized set of value cells' intra-table contexts into a single vector (see Figure~\ref{fig:agg_inter_neighbor}) by leveraging the self-attention mechanism \cite{vaswani2017attention}.
Specifically, we arrange all extracted intra-table contexts of $\mathcal{N}_{v}(t^{m,n}_k)$ into a matrix of $\mathbf{E}^{v}_{t^{m,n}} \in \mathbb{R}^{|\mathcal{N}_{v}(t^{m,n}_k)| \times D_a}$, where each row contains the context aggregated from one value cell of $t_k^{m,n}$. The relative importance for value cells of $t^{m,n}_k$ can be calculated as:
\begin{equation}
	\label{eqn:agg_inter_neighbor_weights}
	\mathbf{\Omega}^{v}_{t^{m,n}} = \text{softmax} \left( \mathbf{W}_s \cdot \left(\mathbf{E}^{v}_{t^{m,n}}\right)^\top \right),
\end{equation}
where $\mathbf{W}_s \in \mathbb{R}^{V \times D_a}$ is a parameter matrix for computing the $V$-view weight matrix $\mathbf{\Omega}^{v}_{t^{m,n}} \in \mathbb{R}^{V \times |\mathcal{N}_{v}(t^{m,n}_k)|}$, the $\text{softmax}$ is applied row-wisely, and $V$ is the number of attention heads setting to 2 in practice. Each row of $\mathbf{\Omega}^{v}_{t^{m,n}}$ reflects one view of the value cells importance distribution. Note if $V=1$, $\mathbf{W}_s$ degenerates into a parameter query vector and the $\text{softmax}$ function can be expanded out similarly to Eqn.~(\ref{eqn:agg_intra_column_weights}).
Then, the value cells aggregated context embedding can be computed as:
\begin{equation}
	\label{eqn:agg_inter_neighbor}
	\mathbf{E}^{v}_{t^{m,n}} = \text{mean} \left( \mathbf{\Omega}^{v}_{t^{m,n}} \cdot \mathbf{E}^{v}_{t^{m,n}} \cdot \mathbf{W}_b \right),
\end{equation}
where $\mathbf{W}_b \in \mathbb{R}^{D_a \times D_b}$ is the parameter matrix for transforming into $D_b$-dim value cells aggregated context, and the final output $\mathbf{E}^{v}_{t^{m,n}} \in \mathbb{R}^{D_b}$ is obtained via a element-wise mean pooling function.

\subsubsection{Position aggregation}
\label{subsubsec:approach_agg_inter_position}
Besides linking tables based on their cell values, the unique grid-like structure of relational tables also grants us a valuable foundation for establishing connections between tables based on the cell's relative position inside the table.
The intuition is that for a subset of relational tables with the same schema, i.e., $\{T_k~|~\phi(k)=u\}$ ($1 \leq u \leq U$), cells at the same position in terms of row index $m$ ($1 \leq m \leq \text{max}(\{S_k^{r}~|~\phi(k)=u \})$) and column index $n$ ($0 \leq n \leq \text{max}(\{S_k^{c}~|~\phi(k)=u \})$) could provide useful contextual information to each other.
For example, suppose there is a collection of identical schema tables describing various music albums, knowing any cell of a song track name (or composer) would reveal other cells of the same position also instantiate the same ``Release'' (or ``People'') type. 
We use $\mathcal{N}_s$ to denote position cells, i.e., $\mathcal{N}_{s}(t_k^{m,n}) \coloneqq \{t_{k'}^{m,n}~\,|\,~\phi(k)=\phi(k') \wedge 0 \leq k' \leq K \wedge k'\neq k \}$.

In general domain, the connections between $t_k^{m,n}$ and position cells $\mathcal{N}_s$ maybe sparse because the number of unique table schema $U$ is also large and comparable to total number of tables $K$.
However, an important practical case is relational tables on semi-structured website which consist of a set of detail pages that each contains information about a particular page topic \cite{lockard2018ceres, lockard2019openceres}.
Typically, the factual information of these tables are automatically populated from an underlying database. When the relational table dataset $\mathcal{D}$ is constructed from such semi-structured websites, there are a large number of helpful inter-table connections to position cells $\mathcal{N}_s$.

Without losing generality, we propose to aggregate from position connections, which potentially is a rich source of the target cell's inter-table contextual information. For generating the position cells aggregated context embedding $\mathbf{e}^{s}_{t^{m,n}} \in \mathbb{R}^{D_s}$, we adpot the similar strategy proposed for aggregating value cells (see Section \ref{subsubsec:approach_agg_inter_neighbor}).
Specifically, we arrange all intra-table contexts of $\mathcal{N}_{s}(t^{m,n}_k)$ into matrix $\mathbf{E}^{s}_{t^{m,n}} \in \mathbb{R}^{|\mathcal{N}_{s}(t^{m,n}_k)| \times D_a}$ and substitute it into Eqn.~(\ref{eqn:agg_inter_neighbor_weights}). The result $\mathbf{\Omega}^{s}_{t^{m,n}}$ is further substituted into Eqn.~(\ref{eqn:agg_inter_neighbor}) for computing $\mathbf{e}^{s}_{t^{m,n}}$.
Note that we truncate the number of position cells when $|\mathcal{N}_{s}|$ is too large according to a sampling budget $b$ in practice to maintain computational tractability.
We will test the effectiveness of our proposed method in both general domain and on a specially constructed dataset from semi-structured websites.

\subsubsection{Topic aggregation}
\label{subsubsec:approach_agg_inter_topic}
Another important type of implicit inter-table connections can be discovered by examining the underlying connectivity between the page topic of target cell and cells of other tables.
Relational Web tables are mainly created for conveying knowledge of relations that are relevant to the page topic. The table cells and topic both refer to relevant real world entities \cite{zhang2017effective}.
It is common to see the topic of certain tables appear as the cell values of other relational tables.
In the example of music album tables, the page topic of album name would appear in other relational tables such as the singer's discography.
So, it is also beneficial for the model to extract contextual information from these topic cells.

We use $\mathcal{N}_{p}$ to denote topic cells containing the same value as the page topic $p_k$ of target cell $t^{m,n}_k$, i.e., $\mathcal{N}_{p}(t^{m,n}_k) \coloneqq \{t^{\tilde{m},\tilde{n}}_{k'}~|~t^{\tilde{m},\tilde{n}}_{k'}=p_k \wedge 0 \leq k' \leq K \wedge k'\neq k\}$.
Similar to the treatment of value cells $\mathcal{N}_v$ and position cells $\mathcal{N}_s$, we first apply the intra-table aggregation function $\text{AGG}_a$ (Section~\ref{subsubsec:approach_agg_intra_func}) to extract $\mathcal{N}_{p}$'s intra-table contexts for constructing $\mathbf{E}^{p}_{t^{m,n}}$, and then generate the topic cells aggregated context embedding $\mathbf{e}^{p}_{t^{m,n}} \in \mathbb{R}^{D_p}$ according to Eqn.~(\ref{eqn:agg_inter_neighbor_weights}) and (\ref{eqn:agg_inter_neighbor}).
Different from the inter-table connections made by $\mathcal{N}_v$ and $\mathcal{N}_s$, which are directly linking to the target cell $t^{m,n}_k$, topic cells $\mathcal{N}_p$ connect to the page topic $p_k$ of $t^{m,n}_k$. Instead of simply using $\mathbf{e}^{p}_{t^{m,n}}$ as the part of the contextual information of $t^{m,n}_k$, we fuse it with $p_k$'s initial embedding $\mathbf{e}_{p_k}$. This can be seen as bringing inter-table contextual information into the embedding of page topic, and the result $\mathbf{e}_{p_k}^p=\mathbf{e}_{p_k}\|\mathbf{e}^{p}_{t^{m,n}}$ is substituted into Eqn.~(\ref{eqn:agg_intra_row_weights}). In this way, we incorporate the global contexts of topic cells into the intra-table aggregation function $\text{AGG}_a$ of the model  (Section~\ref{subsubsec:approach_agg_intra_row}).

By aggregating from column and row cells, as well as various types of inter-table connections to value, position and topic cells, the proposed \textsc{TCN} fully considers both the intra- and inter-table contextual information during learning. Next, we present the fusion of all contexts and the training procedures of the model.

\subsection{Training Objectives}
\label{subsec:approach_obj}
After generating the intra-table contextual embedding $\mathbf{e}^{a}_{t^{m,n}}$ and different inter-table contextual embeddings $\mathbf{e}^{v}_{t^{m,n}}$, $\mathbf{e}^{s}_{t^{m,n}}$, $\mathbf{e}^{p}_{t^{m,n}}$, the final latent representation of target cell $t^{m,n}$ can be computed as:
\begin{equation}
	\label{eqn:agg_fusion}
	\mathbf{h}_{t^{m,n}} = \sigma \left( \mathbf{W}_{h} \cdot \left(\mathbf{e}_{t^{m,n}} \| \mathbf{e}^{a}_{t^{m,n}} \| \mathbf{e}^{v}_{t^{m,n}} \| \mathbf{e}^{s}_{t^{m,n}} \right) \right),
\end{equation}
where $\mathbf{W}_{h}$ is the parameter matrix for fusing the initial cell embedding $\mathbf{e}_{t^{m,n}}$ with all intra- and inter-table contextual embeddings into the final $D_h$-dim representation of $t^{m,n}$.
Note that topic cells aggregated context $\mathbf{e}^{p}_{t^{m,n}}$ is incorporated in $\mathbf{e}^{a}_{t^{m,n}}$ via $\text{AGG}_a$.

\subsubsection{Multi-tasking}
The proposed \textsc{TCN} can be trained in a supervised mode by jointly predicting the type of columns and pairwise relation between columns for each relational table.
Since both of these two multi-class classification tasks are on table column level, we compute the embedding $\mathbf{h}_{t^{*,n}_k} \in \mathbb{R}^{D_h}$ of table column $t^{*,n}_k$ as the mean of its cell embeddings, i.e., $\mathbf{h}_{t^{*,n}_k}=\textsc{Avg} \left( \{ \mathbf{h}_{t^{m,n}_k} \}^{S^r_k}_{m=0} \right)$.
For column type prediction, we use a single dense layer as the final predictive model. The discrepancy between the predicted type distribution and the ground truth column type is measured by the loss function $\mathcal{J}^{\mathcal{C}}$. Specifically, given $c_{t^{*,n}_k} \in \mathcal{C}$ denoted as the true type of $t^{*,n}_k$, we employ the following cross-entropy objective:
\begin{equation}
	\label{eqn:obj_type}
	\scale[0.995]{
	\mathcal{J}^{\mathcal{C}}_k = - \sum^{S^c_k}_{n=0} \sum_{c \in \mathcal{C}} \mathbb{I}_{c_{t^{*,n}_k}=c} \cdot \log \frac{\text{exp}\left( \mathbf{M}_c \cdot \mathbf{h}_{t^{*,n}_k} \right)}{\sum_{c' \in \mathcal{C}} \text{exp}\left( \mathbf{M}_{c'} \cdot \mathbf{h}_{t^{*,n}_k} \right)},
	}
\end{equation}
where $\mathbf{M}_c$ is the parameter matrix for column type $c \in \mathcal{C}$ and $\mathbb{I}$ is an indicator function. 

Similarly, we concatenate the embeddings of a pair of subject and object columns $(t^{*,0}_k, t^{*,n}_k)$ for feeding into a dense layer to generate the prediction on the true relation $r_{t^{*,n}_k}\in \mathcal{R}$ between them:
\begin{equation}
	\label{eqn:obj_relation}
	\scale[0.985]{
	\mathcal{J}^{\mathcal{R}}_k = - \sum^{S^c_k}_{n=1} \sum_{r \in \mathcal{R}} \mathbb{I}_{r_{t^{*,n}_k}=r} \cdot \log \frac{\text{exp}\left( \mathbf{M}_r \cdot \left(\mathbf{h}_{t^{*,0}_k} \,\|\, \mathbf{h}_{t^{*,n}_k} \right) \right)}{\sum_{r' \in \mathcal{R}} \text{exp}\left( \mathbf{M}_{r'} \cdot \left(\mathbf{h}_{t^{*,0}_k} \,\|\, \mathbf{h}_{t^{*,n}_k} \right) \right)},
	}
\end{equation}
where $\mathbf{M}_r$ is parameter matrix for pairwise column relation $r \in \mathcal{R}$.

So, given a mini-batch of relational tables $\mathcal{B} \subseteq \mathcal{D}$, the overall training objective $\mathcal{J}$ of the proposed \textsc{TCN} can be obtained via a convex combination of the above two tasks' loss functions:
\begin{equation}
	\label{eqn:obj_overall}
	\mathcal{J} = \sum_{k \in \mathcal{B}} \gamma\mathcal{J}^{\mathcal{C}}_k + (1-\gamma)\mathcal{J}^{\mathcal{R}}_k,
\end{equation}
where $\gamma$ is a mixture hyperparameter for balancing the magnitude of two objectives for predicting column type and pairwise relation.

\subsubsection{Unsupervised pre-training}
\label{subsubsec:approach_training_pretraining}
Learning relational table representation directly under the supervision of column type and relation labels are not always feasible due to the expensive cost for obtaining high quality annotations.
The proposed \textsc{TCN} can also be trained in an unsupervised way without relying on explicit labels.
Specifically, we first train \textsc{TCN} according to the pre-training objective to obtain the output cell embeddings. We then use these pre-trained cell embeddings as initialization for the supervised fine-tuning phase aimed at jointly predicting column type and pairwise column relation (Eqn.~(\ref{eqn:obj_overall})).
Similar to the the Masked Language Model (MLM) objective of \textsc{BERT} \cite{devlin2018bert}, we randomly mask $10\%$ of table cells beforehand for recovery. Given a masked cell $\hat{t}^{m,n}_k$ and the global context-aware embedding $\mathbf{h}_{\hat{t}^{m,n}_k}$ learned by \textsc{TCN}, the objective for predicting the original cell value is computed as:
\begin{equation}
	\label{eqn:obj_pretraining}
	\scale[0.95]{
	\mathcal{J} = - \sum_{k \in \mathcal{B}} \sum_{\hat{t}^{m,n}_k \in \hat{T}_k} \sum_{v \in \mathcal{V}} \mathbb{I}_{t^{m,n}_k=v} \cdot \log \frac{\text{exp}\left( \mathbf{M}_v \cdot \mathbf{h}_{\hat{t}^{m,n}_k} \right)}{\sum_{v' \in \mathcal{V}} \text{exp}\left( \mathbf{M}_{v'} \cdot \mathbf{h}_{\hat{t}^{m,n}_k} \right)},
	}
\end{equation}
where $\hat{T}_k$ is the set of all masked cells of the $k$-th table, $\mathcal{V}$ is the set of all cell values of $\mathcal{D}$, and $\mathbf{M}_v$ is parameter matrix for cell value $v$ which could include one or multiple words after the normalization (Section~\ref{subsubsec:approach_agg_inter_neighbor}).
The pre-trained cell embeddings can later be used for initializing fine-turning phase. As we show in experiments (Section~\ref{subsec:experiments_pretraining}), pre-training improves the performance by $+2.0\%$ and $+2.3\%$ of F1-weighted for predicting column type and relation.

\subsection{Complexity}
Assuming the attention-based intra-table aggregation function $\text{AGG}_a$ takes constant time, the per-batch time complexity of the proposed TCN is $\mathcal{O}(|B|\,(b_v + b_s + b_p))$ in principle, where $|B|$ is batch size and $b_v$ (and $b_s$, $b_p$) is the sampling budget for inter-table connections of value (and position, topic) cells.
In practice, we simply set $b_v = b_s = b_p$ so the time complexity becomes $\mathcal{O}(|B|b)$ which is linear to the product of batch size $|B|$ and sampling budget $b$.
In optimized implementation, we can process tables in batch into one large table beforehand by padding and concatenating cells which can further reduce the complexity to $\mathcal{O}(b)$. This allows us to scale the model to tens of millions tables while remaining control on the balance between model expressiveness and efficiency.

\section{Experiments}
\label{sec:experiments}

\begin{table}[t]
	\renewcommand{\arraystretch}{1.2}
	\centering
	\caption{Statistics of two real Web table datasets $\mathcal{D}^{m}$ and $\mathcal{D}^{w}$.}
	\label{tab:datasets}
	\vspace{-0.1in}
	\scale[0.9]{
	\begin{tabular}{|l||c|c|c|c|c|}
	\hline
		 & {\textbf{\#tables}} & \textbf{Avg. \#} & \textbf{Avg. \#} & \textbf{\#column} & \textbf{\#pairwise} \\ 
		 & {\textbf{K}} & \textbf{rows $S_k^r$} & \textbf{cols. $S_k^c$} & \textbf{types $|\mathcal{C}|$} & \textbf{relations $|\mathcal{R}|$} \\ \hline \hline
		$\mathcal{D}^{m}$ & {128,443} & {7.0} & {3.6} & {8} & {14} \\ \hline
		$\mathcal{D}^{w}$ & {55,318} & {16.1} & {2.4} & {201} & {121} \\ \hline \hline
		 & {\textbf{\#schemas}} & \textbf{Avg. \#value} & \multicolumn{2}{|c|}{\textbf{Avg. \#position}} & \textbf{Avg. \#topic} \\
		 & {\textbf{U}} & \textbf{cells $|\mathcal{N}_v|$} & \multicolumn{2}{|c|}{\textbf{cells $|\mathcal{N}_s|$}} & \textbf{cells $|\mathcal{N}_p|$} \\ \hline \hline
		$\mathcal{D}^{m}$ & {11} & {10.7} & \multicolumn{2}{|c|}{5048.1} & {2.4} \\ \hline
		$\mathcal{D}^{w}$ & {6,538} & {7.2} & \multicolumn{2}{|c|}{0.9} & {1.9} \\ \hline
	\end{tabular}
	}
	\vspace{-0.1in}
\end{table}

\begin{table*}[t]
	\renewcommand{\arraystretch}{1.25}
	\centering
	\caption{Performance of baselines and variants of \textsc{TCN} and on predicting column type $\mathcal{C}$ and pairwise column relation $\mathcal{R}$ on dataset $\mathcal{D}^{m}$. For all metrics, higher values indicate better performance. Bold highlights global highest values. Underline denotes best performance among baselines. Relative improvements over the base variant \textsc{TCN}-intra are shown in parenthesis.}
	\label{tab:results_overall}
	\vspace{-0.1in}
	\scale[0.975]{
	\begin{tabular}{|l||c|c|c|c|c|c|}
	\hline
		\multirow{2}*{\textbf{Method}} & \multicolumn{3}{|c|}{\textbf{Column type} $\mathcal{C}$} & \multicolumn{3}{|c|}{\textbf{Pairwise column relation} $\mathcal{R}$} \\ \cline{2-7}
		 & {Acc.} & {F1-weighted} & {Cohen's kappa $\kappa$} & {Acc.} & {F1-weighted} & {Cohen's kappa $\kappa$} \\ \hline \hline
		{\textsc{Table2Vec}}		& {.832} & {.820} & {.763} & {.822} & {.810} & {.772} \\ 
		{\textsc{TaBERT}}		& {.908} & {.861} & {.834} & {.877} & {.870} & {\underline{.846}} \\
		{\textsc{TURL}}			& {.914} & {.877} & {\underline{.876}} & {\underline{.890}} & {\underline{.889}} & {.838} \\ \hline
		{\textsc{HNN}}			& {.916} & {.883} & {.869} & {.848} & {.843} & {.794} \\
		{\textsc{Sherlock}}		& {\underline{.922}} & {\underline{.895}} & {.863} & {.831} & {.818} & {.802} \\ \hline \hline
		{\textsc{TCN}-intra}					& {.911} & {.881} & {.873} & {.893} & {.894} & {.869} \\ 
		{\textsc{TCN}-$\mathcal{N}_v$}		& {{.939 (+3.1\%)}} & {{.916 (+4.0\%)}} & {{.897 (+2.8\%)}} & {{.920 (+3.0\%)}} & {.920 (+2.9\%)} & {{.898 (+3.3\%)}} \\ 
		{\textsc{TCN}-$\mathcal{N}_s$}		& {.934 (+2.5\%)} & {.908 (+3.1\%)} & {.894 (+2.4\%)} & {.908 (+1.7\%)} & {.912 (+2.0\%)} & {.881 (+1.4\%)} \\
		{\textsc{TCN}-$\mathcal{N}_p$}		& {.923 (+1.3\%)} & {.890 (+1.0\%)} & {.880 (+0.8\%)} & {.906 (+1.4\%)} & {{.904 (+1.1\%)}} & {{.875 (+0.7\%)}} \\ \hline
		{\textsc{TCN}}							& {\textbf{.958 (+5.2\%)}} & {\textbf{.938 (+6.5\%)}} & {\textbf{.913 (+4.6\%)}} & {\textbf{.934 (+4.6\%)}} & {\textbf{.925 (+3.5\%)}} & {\textbf{.905 (+4.1\%)}} \\ \hline
	\end{tabular}
	}
	\vspace{-0.15in}
\end{table*}

In this section, we evaluate the performance of the proposed \textsc{TCN} for relational table representation learning against competitive baselines on two real world large-scale relational Web tables datasets.
Particularly, we aim at answering the following research questions:

\begin{itemize}
	\item \textbf{RQ1}: How does the proposed method perform compared with the state-of-the-art methods for predicting relational table's column type and pairwise relation between columns?
	\item \textbf{RQ2}: How does each type of the proposed inter-table aggregation module affect the overall performance of the model?
	\item \textbf{RQ3}: Is the proposed method also effective when applied for pre-training on unlabeled corpus of relational tables?
	\item \textbf{RQ4}: What are some concrete examples of the column type and column pairwise relations discovered by the model?
	\item \textbf{RQ5}: What are the recommended hyper-parameter settings for applying the proposed model in practical cases?
\end{itemize}

\subsection{Datasets}
We collected a dataset $\mathcal{D}^{m}$ of 128K relational Web tables from 6 mainstream semi-structured websites in the music domain. Tables of $\mathcal{D}^{m}$ generally fall into three page topic categories: ``Person'' (e.g., singer, composer and etc.), ``Release'' (i.e., music album), and ``Recording'' (i.e., song track). The number of unique table schemas in $\mathcal{D}^{m}$ is relatively small ($U=11$) because tables are automatically populated. We manually annotated each table schema to obtain column type and pairwise column relation information.
For relational Web tables in the general domain, we utilize datasets provided by Deng \textit{et al.} \cite{deng2020turl} containing annotated relational tables from a raw corpus of 570K Web tables on Wikipedia. We build a dataset $\mathcal{D}^{w}$ by taking a subset of 5.5K tables with annotations on both column type and relation. Specifically, we take the intersection of task-specific datasets for column type annotation and relation extraction described in the paper.
For both datasets, we keep tables with at least 2 columns/rows.
More descriptive statistics are provided in Table \ref{tab:datasets}.

\subsection{Experimental Settings}
\subsubsection{Baseline methods.}
We compare the proposed \textsc{TCN} against the state-of-the-art tabular data representation learning methods:
\begin{itemize}
	\item \textsc{Table2Vec} \cite{zhang2019table2vec}: This method flattens a cropped portion of relational table and its topic into a sequence of cell tokens and uses \textsc{word2vec} \cite{mikolov2013distributed} for learning column/cell embeddings.
	\item \textsc{TaBERT} \cite{yin2020tabert}: It jointly learns embeddings of natural language sentences and relational tables using its content snapshots to carve out relevant rows for feeding into \textsc{BERT} \cite{devlin2018bert}.
	\item \textsc{TURL} \cite{deng2020turl}: This method uses linearized cells and linked entities as input into its \text{Transformer} \cite{vaswani2017attention} based encoder enhanced with structure-aware visibility matrix for pre-training.
\end{itemize}
Besides, we consider methods that are specifically designed for column type prediction or pairwise column relation extraction:
\begin{itemize}
	\item \textsc{HNN} \cite{chen2019learning}: This method models the contextual semantics of relational table column using a bidirectional-RNN with an attention layer and learns column embeddings with a CNN.
	\item \textsc{Sherlock} \cite{hulsebos2019sherlock}: It utilizes four categories of statistical features describing the character/word distributions and semantic features to predict the semantic type on column level.
\end{itemize}
We use open-source implementations provided by the original papers for all baseline methods and follow the recommended setup guidelines when possible.
For \textsc{TaBERT}, we use the page topic as its NL utterance. And we set the embedding of linked entity in \textsc{TURL} the same as its cell since they are not provided as input in our case.
As \textsc{HNN} and \textsc{Sherlock} are originally designed for predicting semantic type of columns, we concatenate the embeddings of subject and object pair of columns to predict the relation.
We also tested with representative methods (e.g., \textsc{MTab}, \textsc{MantisTable}, \textsc{CVS2KG}) from the SemTab 2019 challenges\footnote{http://www.cs.ox.ac.uk/isg/challenges/sem-tab/2019/}. But we were only able to successfully generate results from \textsc{MTab} due to the high time complexity of others. We observed a relative large performance gap between it and \textsc{Table2Vec}'s (probably because of its dependence on pre-linked entities) so we exclude it in following discussions.
For all methods, we use the same random split of 80/10/10 percents of tables for training/validation/test at each round and we report the average performance of five runs.
For \textsc{TCN}, the sampling budget $b$ for inter-table connections is set to 20 and objective mixture weight $\gamma$ is set to 0.5.
We set the dimension of cell and all context embeddings as $300$ and initialize each cell embedding by matching with \textsc{Glove} embeddings \cite{pennington2014glove} (taking mean in case of multiple tokens).

\subsubsection{Evaluation metrics.} For classifying multi-class column type $\mathcal{C}$ and pairwise column relation $\mathcal{R}$, we use metrics of \textit{mean accuracy} (Acc.), F1-weighted score and the \textit{Cohen's kappa} $\kappa$ coefficient.

\begin{figure}[t]
    \centering
    \subfigure[Performance of \textsc{TCN} and baselines on dataset $\mathcal{D}^{w}$ in terms of accuracy.]
    {\includegraphics[width=1\linewidth]{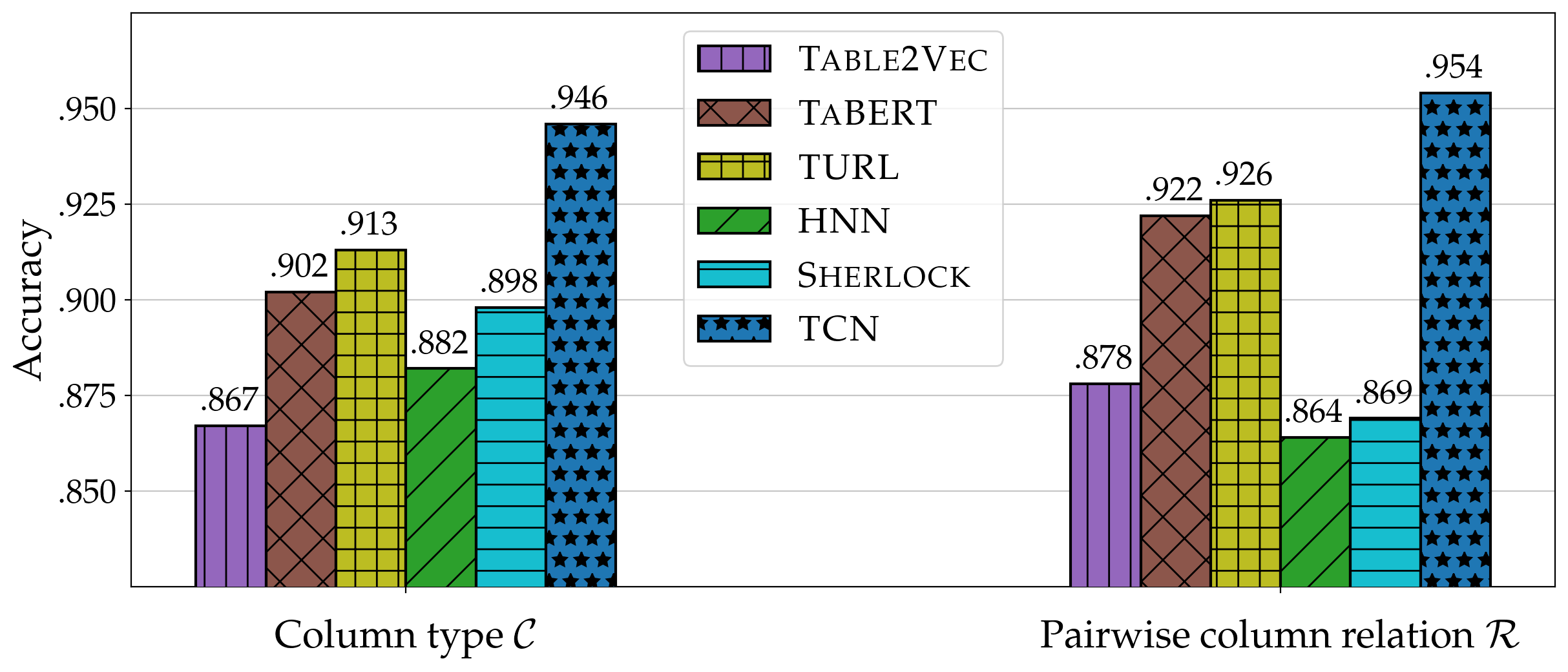}\label{fig:performance_dw_acc}}
	\vspace{-0.1in}
    \subfigure[Performance of \textsc{TCN} and baselines on dataset $\mathcal{D}^{w}$ in terms of F1-weighted.]
    {\includegraphics[width=1\linewidth]{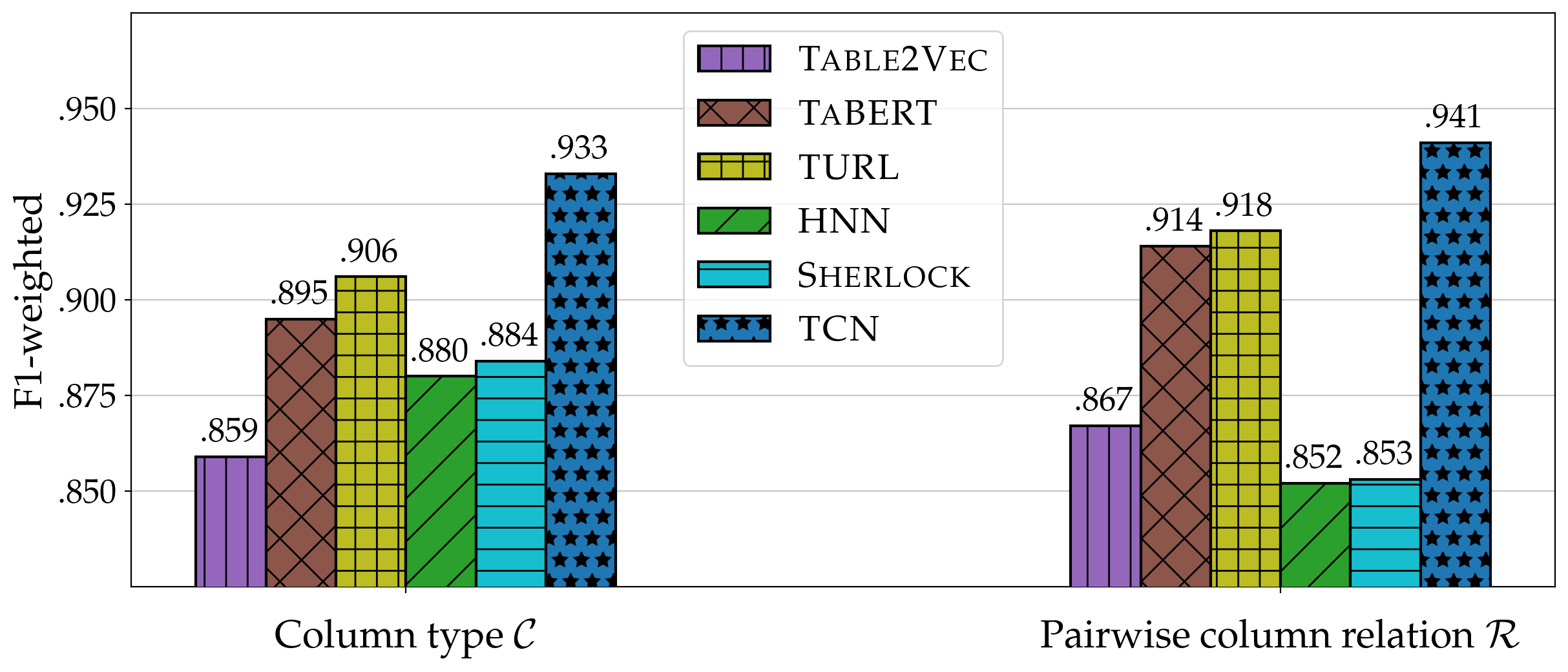}\label{fig:performance_dw_f1}}
    \caption{The proposed \textsc{TCN} can consistently outperform baseline methods for column type $\mathcal{C}$ prediction and pairwise column relation $\mathcal{R}$ extraction on open domain dataset $\mathcal{D}^{w}$.}
    \label{fig:performance_dw}
    \vspace{-0.15in}
\end{figure}

\subsection{Overall Performance (RQ1)}
Table~\ref{tab:results_overall} and Figure~\ref{fig:performance_dw} presents the experimental results of applying the proposed \textsc{TCN} and baseline methods on predicting relational Web table column type $\mathcal{C}$ and pairwise column relation $\mathcal{R}$ on dataset $\mathcal{D}^{m}$ and $\mathcal{D}^{w}$ respectively. In the following discussions, we refer to F1-weighted as F1 unless explicitly stated otherwise.

For tabular data representation learning methods, \textsc{TURL} performs better than other baselines on both tasks across datasets.
\textsc{Table2Vec} underperforms all other methods because it simply crops and flattens part of the table for feeding into the shallow \textsc{word2vec}.
There is quite large performance margin ($+5.0\%$ and $+7.4\%$ for two tasks on $\mathcal{D}^{m}$ in terms of F1) from \textsc{Table2Vec} to \textsc{TaBERT} using a deep \textsc{BERT}-based encoder.
\textsc{TURL} enhances the deep encoder with visibility mask which can partially capture the grid-like structure of relational table so it has better performance compared with \textsc{Table2Vec} and \textsc{TaBERT}.
This means better modeling the intra-table context is beneficial. But these methods all rely on linearizing the table into a long sequence of cell tokens.

For methods specifically designed for column type prediction, \textsc{Sherlock} performs better than \textsc{HNN}
for predicting column type $\mathcal{C}$ on two datasets. However, it cannot produce competitive results for predicting pairwise column relation $\mathcal{R}$
 on two datasets which is roughly on the same performance level as \textsc{Table2Vec}. It is interesting to note \textsc{Sherlock} achieves the best performance for predicting column type $\mathcal{C}$ on $\mathcal{D}^{m}$ among all baselines because of the effectiveness of its statistical features. But none of these baselines is capable of modeling the valuable inter-table contextual information.

The proposed \textsc{TCN} can consistently outperform baseline methods on all metrics across two datasets. For predicting the column type $\mathcal{C}$, \textsc{TCN} scores an F1 of .938 on $\mathcal{D}^{m}$ which is $+4.8\%$ relatively over \textsc{Sherlock} ($+7.0\%$ relatively over \textsc{TURL}), and scores an F1 of .933 on $\mathcal{D}^{w}$ which is $+5.5\%$ relatively over \textsc{Sherlock} ($+3.0\%$ relatively over \textsc{TURL}). For predicting the pairwise column relation $\mathcal{R}$, \textsc{TCN} can generate an F1 of .925 on $\mathcal{D}^{m}$ which is $+4.1\%$ relatively over \textsc{TURL}, and score an F1 of .941 on $\mathcal{D}^{w}$ which is $+2.5\%$ relatively over \textsc{TURL}. This justifies the effectiveness \textsc{TCN}'s inter-table aggregation modules for capturing the contextual information across various types of implicitly connected tables. These inter-table contexts are complementary to the intra-table context providing additional discriminative power in downstream tasks.

\begin{figure}[t]
    \centering
    \subfigure[Column type $\mathcal{C}$]
    {\includegraphics[width=0.475\linewidth,]{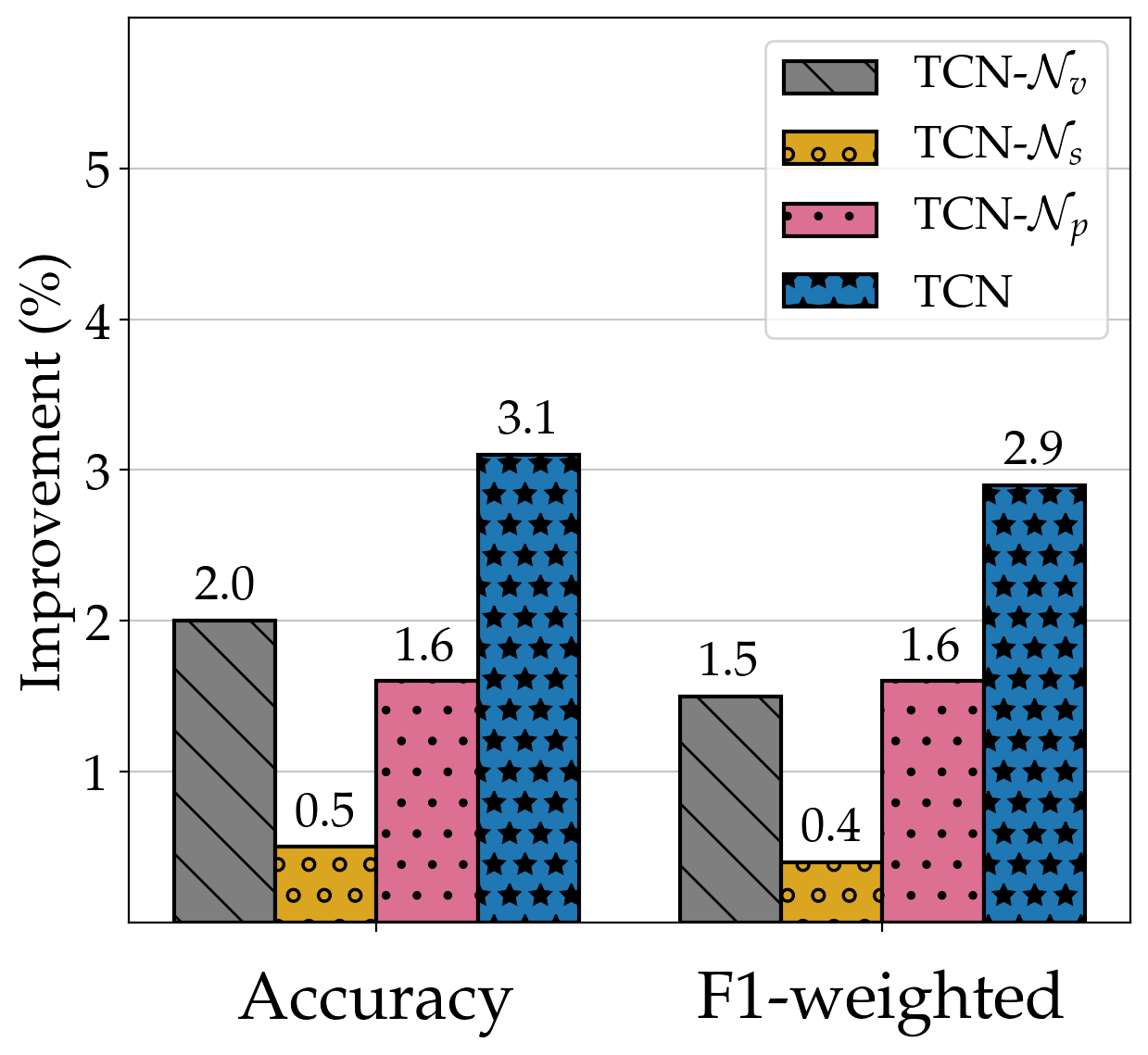}\label{fig:performance_dw_acc}}
	\hfill
    \subfigure[Pairwise column relation $\mathcal{R}$]
    {\includegraphics[width=0.475\linewidth]{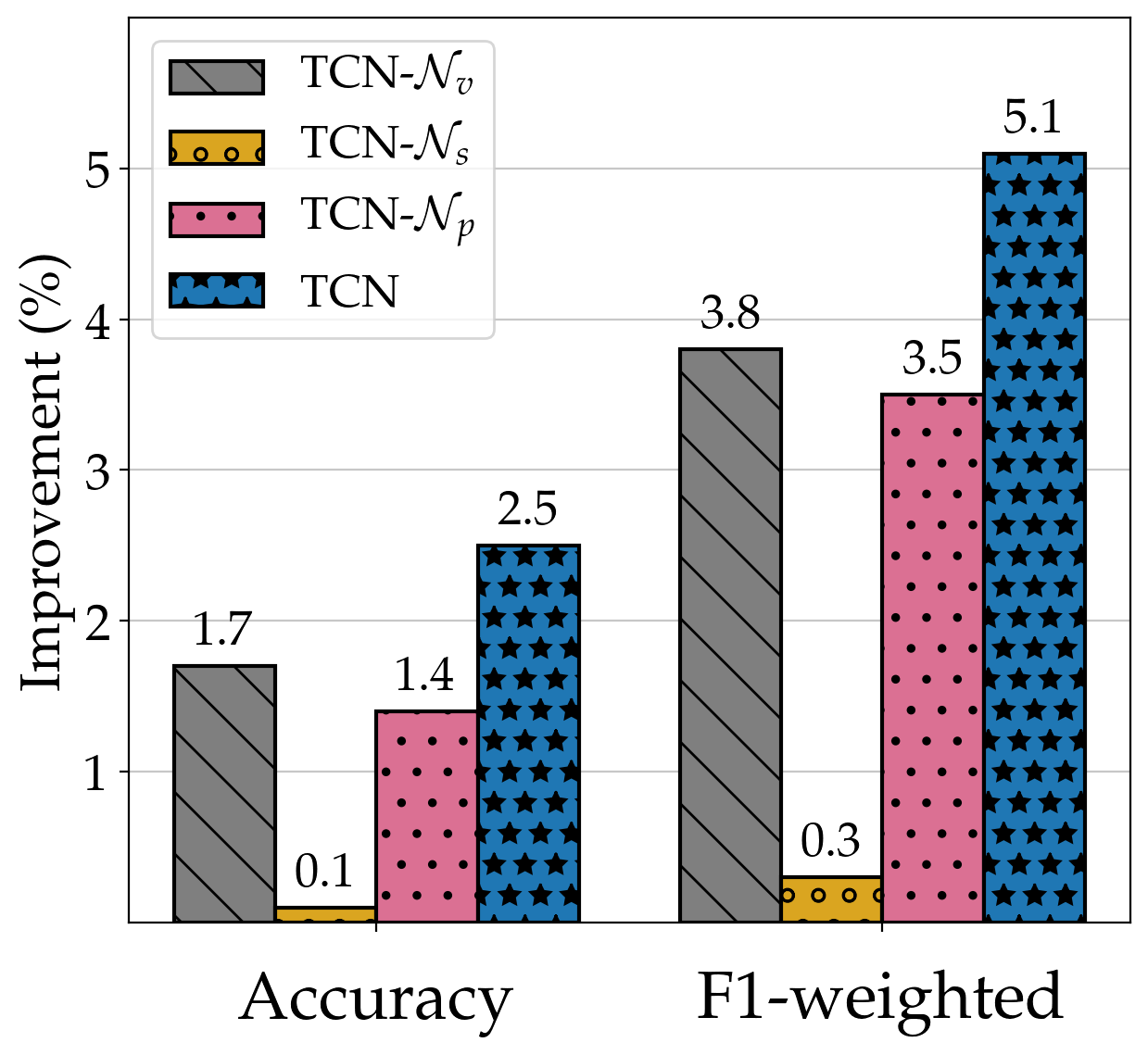}\label{fig:performance_dw_f1}}
	\vspace{-0.1in}
    \caption{Relative improvements of \textsc{TCN} and variants over the base variant \textsc{TCN}-intra for predicting column type $\mathcal{C}$ and pairwise column relation $\mathcal{R}$ on the open domain dataset $\mathcal{D}^{w}$.}
    \label{fig:results_ablation_relation}
    \vspace{-0.1in}
\end{figure}

\subsection{Ablation Studies (RQ2)}
To further validate the effectiveness of each inter-table aggregation module of \textsc{TCN}, we propose 4 model variants by including selected type(s) of inter-table aggregations and compare against them:
\begin{itemize}
	\item \textsc{TCN}-intra: Base version only considers intra-table contexts, i.e., cells of the same column/row, using the intra-table aggregation $\text{AGG}_a$ (Section~\ref{subsubsec:approach_agg_intra_func}) on each table independently.
	\item \textsc{TCN}-$\mathcal{N}_v$: This variant considers the inter-table context of value cells $\mathcal{N}_v$ (Section~\ref{subsubsec:approach_agg_inter_neighbor}) besides intra-table context.
	\item \textsc{TCN}-$\mathcal{N}_s$: Position cells $\mathcal{N}_s$ of the same schema position (Section~\ref{subsubsec:approach_agg_inter_position}) are considered as the inter-table context here.
	\item \textsc{TCN}-$\mathcal{N}_p$: Topic cells $\mathcal{N}_p$ of the same value as target cell's page topic (Section~\ref{subsubsec:approach_agg_inter_topic}) are considered in this variant.
\end{itemize}
The performance of \textsc{TCN} and its variants are presented in Table~\ref{tab:results_overall} and Figure~\ref{fig:results_ablation_relation}.
The base variant \textsc{TCN}-intra which only considers intra-table contextual information performs roughly on the same level as the best baseline model \textsc{TURL} for tabular data representation learning.
This demonstrates the intra-table aggregation function $\text{AGG}_a$ can successfully summarize the contextual information inside the target table from cells of the same column and row.
We compare other \textsc{TCN}'s variant against \textsc{TCN}-intra for evaluating the relative contribution of each inter-table aggregation module.

The \textsc{TCN}-$\mathcal{N}_v$ considering inter-table contexts from value cells provides a significant increase on the performance for both tasks.
By only aggregating from value cells, \textsc{TCN}-$\mathcal{N}_v$ can score F1s of .916 and .921 for predicting $\mathcal{C}$ on two datasets which are relatively $+2.4\%$ and $+3.6\%$ over the baseline \textsc{Sherlock} (similar trend also hold for predicting $\mathcal{R}$).
This indicates value cells of the same value as target cell consistently provide rich inter-table contextual information complementary to the intra-table contexts, and the value cells aggregation module of \textsc{TCN} is effective in learning from them.

The \textsc{TCN}-$\mathcal{N}_s$ considering position cells besides the intra-table context gives high performance for both tasks on dataset $\mathcal{D}^{m}$ but is less helpful on dataset $\mathcal{D}^{w}$. 
This is because of the difference in two datasets: $\mathcal{D}^{m}$ contains relational tables from semi-structured websites with rich connections between position cells ($\overline{|\mathcal{N}_s|}>5 \times 10^3$) while $\mathcal{D}^{w}$ is constructed from open domain Wikipedia pages with sparse position cells ($\overline{|\mathcal{N}_s|}=0.9$). Given relatively densely connected relational table schemas, we found aggregating from position cells can provide us useful inter-table contextual information.

The \textsc{TCN}-$\mathcal{N}_p$ considering topic cells besides the intra-table context generally gives good improvements of performance for both tasks on two datasets 
comparable to the improvements provided by \textsc{TCN}-$\mathcal{N}_v$.
This confirms that aggregating from topic cells of other tables can provide \textsc{TCN}-$\mathcal{N}_p$ additional contextual information. And, at last, by considering all three types of inter-table aggregation modules, the full version of \textsc{TCN} can consistently improve upon \textsc{TCN}-intra which focuses only on the intra-table context by $+6.5\%$ and $+5.1\%$ for two tasks across datasets in terms of F1.

\begin{table}[t]
	\renewcommand{\arraystretch}{1.2}
	\centering
	\caption{Performance of \textsc{TCN} and baselines under different training settings evaluated in terms of F1-weighted.}
	\label{tab:results_pretraining}
	\vspace{-0.05in}
	\scale[0.975]{
	\begin{tabular}{|l||c|c|c|c|c|}
	\hline
		\multirow{2}*{\textbf{Method}} & \multicolumn{2}{|c|}{$\mathcal{D}^{m}$} & \multicolumn{2}{|c|}{$\mathcal{D}^{w}$} \\ \cline{2-5}
		 & {Type $\mathcal{C}$} & {Relation $\mathcal{R}$} & {Type $\mathcal{C}$} & {Relation $\mathcal{R}$} \\ \hline \hline
		{\textsc{TCN} (supervised)}			& {.938} & {.925} & {.933} & {.941} \\ \hline
		{\textsc{TCN} + \textsc{TaBERT}}	& \multirow{2}*{.948} & \multirow{2}*{.933} & \multirow{2}*{.942} & \multirow{2}*{.949} \\ 
		{for pre-training} 					& & & & \\ \hline
		{\textsc{TCN} + \textsc{TURL}}		& \multirow{2}*{.945} & \multirow{2}*{.934} & \multirow{2}*{.946} & \multirow{2}*{.953} \\ 
		{for pre-training} 					& & & & \\ \hline
		{\textsc{TCN} full w/}				& \multirow{3}*{.957} & \multirow{3}*{.946} & \multirow{3}*{.951} & \multirow{3}*{.960} \\ 
		{pre-training} 						& & & & \\
		{\& fine-tuning} 					& & & & \\ \hline
	\end{tabular}
	}
	\vspace{-0.15in}
\end{table}

\subsection{Unsupervised Pre-training (RQ3)}
\label{subsec:experiments_pretraining}
Besides training the proposed \textsc{TCN} under the explicit supervision of column type and relation labels, we can also pre-train \textsc{TCN} on large-scale unlabeled corpus of relational Web tables and fine-tune on downstream tasks (see Section \ref{subsubsec:approach_training_pretraining}). 
We also leverage two baseline methods (\textsc{TaBERT} and \textsc{TURL}) capable of pre-training on unlabeled relational tables to obtain initial cell embeddings as input for supervised multi-task training of \textsc{TCN}. We conduct comparative experiments on both datasets and present the results in Table~\ref{tab:results_pretraining}.

We can see that incorporating the pre-trained cell embeddings of two baseline methods can improve the final performance of \textsc{TCN} on both tasks of predicting column type $\mathcal{C}$ and pairwise column relation $\mathcal{R}$.
The performance improvement gained by utilizing the pre-trained embeddings of \textsc{TURL} is similar to \textsc{TaBERT} since they both focus on effectively summarizing intra-table contextual information during the pre-training phase.
In contrast to \textsc{TaBERT} and \textsc{TURL}, the proposed \textsc{TCN} can naturally incorporate different types of inter-table context during the pre-training phase without column type or relation labels. So, by combing the intra- and inter-table contexts learned during both the pre-training and fine-tuning phases, the propose \textsc{TCN} can score F1s of .957 and .951 for predicting $\mathcal{C}$ on two datasets ($+2.0\%$ and $+1.9\%$ relatively over \textsc{TCN} without pre-training); and scores F1s of .946 and .960 for predicting $\mathcal{R}$ ($+2.3\%$ and $+2.0\%$ relatively over \textsc{TCN} without pre-training).

\subsection{Qualitative Analysis (RQ4)}
Our goal of \textsc{TCN} is to automatically discover column type and pairwise relation between columns given a relational Web table and an external knowledge base. We provide concrete examples on the output of \textsc{TCN} to show its effectiveness  and practical utilities.

We used a proprietary music ontology for dataset $\mathcal{D}^{m}$ in experiments which includes column types such as ``Release'', ``People'', ``Recording'', ``RecordLabel'', ``XMLSchema\#string'', and etc.
We found \textsc{TCN} can achieve almost perfect performance on major types (e.g., column [``Lucky Old Sun'', ``Life on a Rock'', ``Cosmic Hallelujah'',...] predicted as ``Release'', and column [``Johnny Cash'', ``Joan Jett'', ``John Keawe'',...] predicted as type ``People'').
We also found that column like [``US'', ``Japan'', ``UK'',...] was classified as the ``XMLSchema\#string'' because there is no corresponding type, i.e., ``Nation'' in the given knowledge base.
This can be optimized by replacing alternative source ontology in the music domain with more complete coverage and finer granularity.
For the task of predicting relations between column pairs, the pre-defined column pairwise relations include ``hasPerformer'', ``hasOriginalReleaseDate'', ``hasDuration'', ``isRecordingOfTrack'', ``isTrackOfRelease'', and etc.
One interesting example of the identified relation between columns of [``Miles Wintner'', ``Harmony Tividad'', ``Sariah Mae'',...] and [``drums'', ``bass'', ``keyboards'',...] is ``playsInstrument'' although the latter column is identified as type ``XMLSchema\#string''.
This indicates the potential advantage of jointly modeling the type and relation of columns to capture their complementary information.

\begin{figure}[t]
    \centering
    \subfigure[Improving the value of sampling budget $b$ is generally beneficial until 20.]
    {\includegraphics[width=0.475\linewidth]{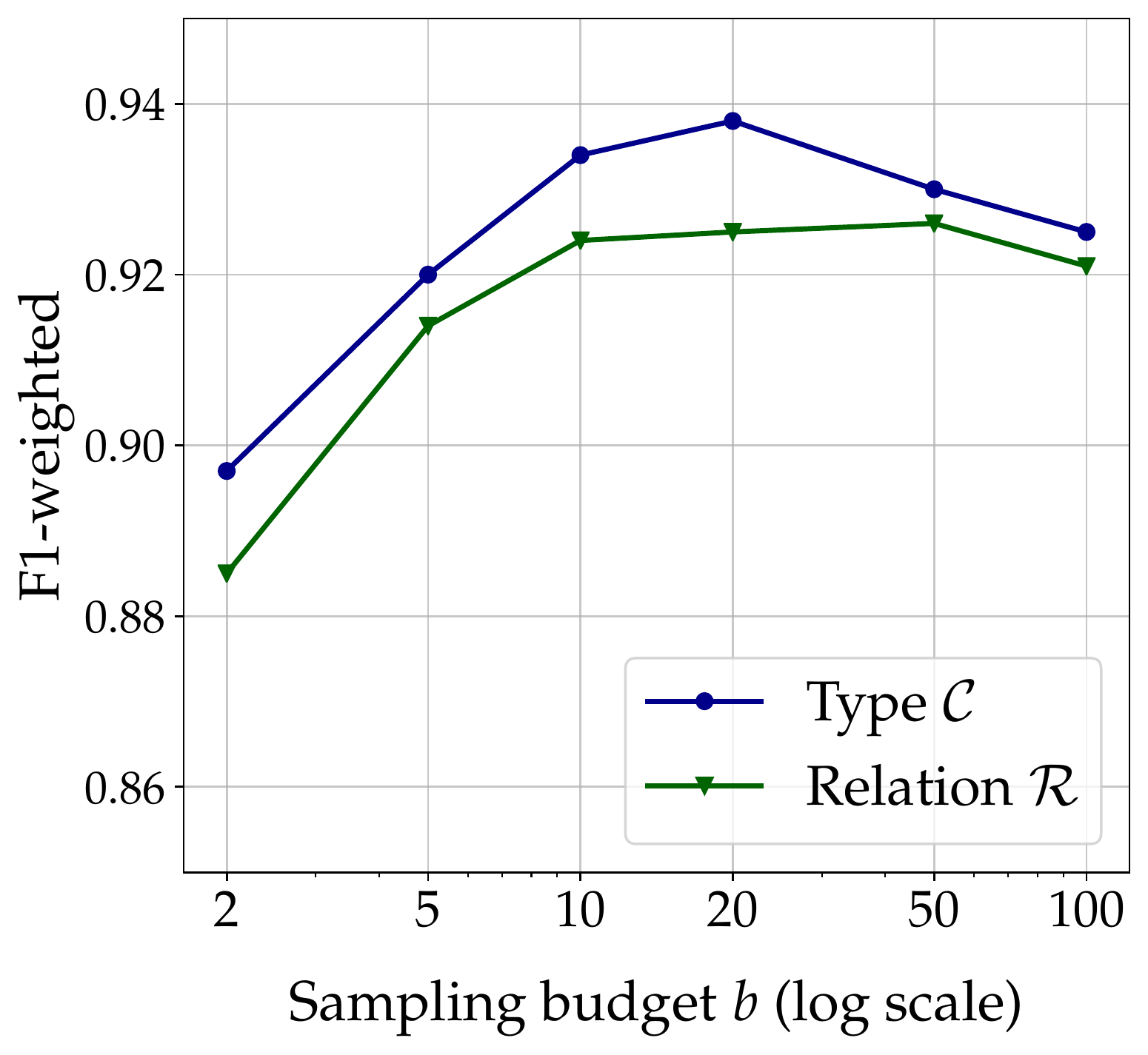}\label{fig:results_sensitivity_b}}
    \hfill
    \subfigure[A choice of $\gamma$ in range of {[0.2, 0.8]} can yield stable model performance.]
    {\includegraphics[width=0.475\linewidth]{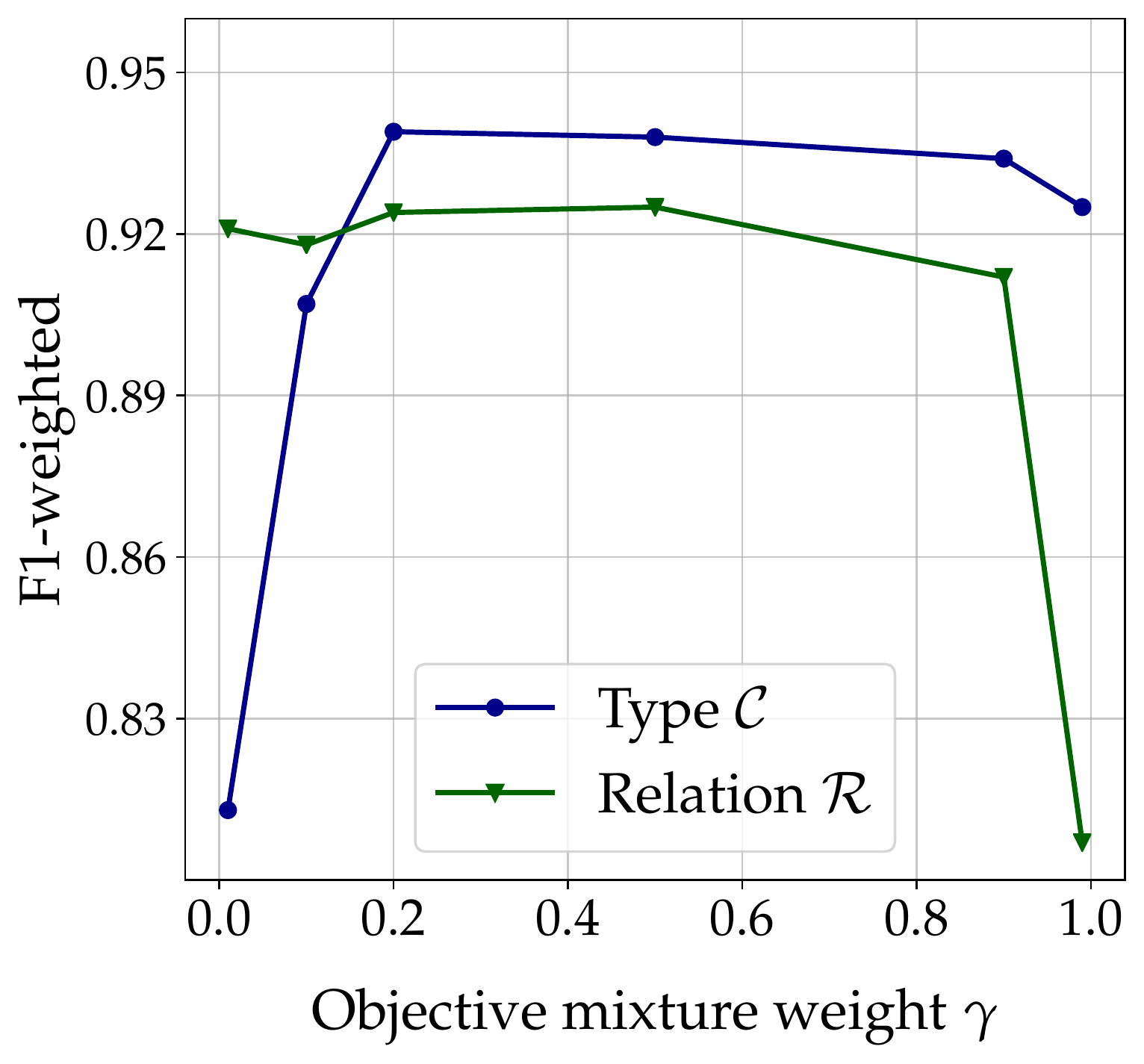}\label{fig:results_sensitivity_gamma}}
	\vspace{-0.1in}
    \caption{Sensitivity of \textsc{TCN}'s performance on different values of sampling budget $b$ and objective mixture weight $\gamma$.}
    \label{fig:results_sensitivity}
    \vspace{-0.15in}
\end{figure}

\subsection{Sensitivity (RQ5)}
We examine the impact of \textsc{TCN}'s key hyper-parameters: (1) the sampling budget $b$ for different types of inter-table connections ($|\mathcal{N}_{b}|$, $|\mathcal{N}_{s}|$, and $|\mathcal{N}_{p}|$), and (2) the mixture weight $\gamma$ of overall objective (Eqn.~(\ref{eqn:obj_overall})), on the model's performance using dataset $\mathcal{D}^{m}$. 
We analyze these two hyper-parameters adopting the gird-search strategy: $b$ is chosen from $\{2, 5, 10, 20, 50, 100\}$ and $\gamma$ is chosen from $\{0.01, 0.1, 0.2, 0.5, 0.9, 0.99\}$.
Figure~\ref{fig:results_sensitivity} presents the results.

In Figure~\ref{fig:results_sensitivity_b} we can see that improving the value of $b$ from 2 to 10 can noticeably improve the performance because larger $b$ values allow \textsc{TCN} to learn from more inter-table connections and thus aggregating more contextual information across tables. But further improving $b$ brings in diminishing benefits and large values such as 100 in turn hurts the performance because of additional noise. In Figure~\ref{fig:results_sensitivity_gamma}, we can see that \textsc{TCN}'s performance is generally stable for a $\gamma$ values in range of {[0.2, 0.8]}. For extreme cases of small (or large) values of $\gamma$, the supervised multi-task objective of \textsc{TCN} will degenerates into single-task setting because the gradients from one task are being squashed leading to the model only focuses on either the column type or pairwise relation prediction.
In practice, we recommend finding the optimal values of $b$ based on the dataset and choosing a balancing value of $\gamma$ such as 0.5.

\section{Conclusions}
\label{sec:conclusions}
 In this work, we proposed a novel approach for learning relational table latent representations. Our proposed method aggregates cells of the same column and row into the intra-table context. In addition, three types of inter-table contexts are aggregated from value cells of the same value, position cells of the same position, and topic cells of the same value as target cell's page topic. Extensive experiments on two real relational table datasets from open domain and semi-structured websites demonstrated the effectiveness of our model.
We focus on relational Web tables of horizontal format in this work but a huge number of tables with various types and structures can be found on the Web.
We consider handling complex table formats such as different orientations, composite header cells, changing subject column index and nested cells as an interesting future direction.
Also, applying the leaned cell embeddings for other downstream tasks such as cell entity linking, table type prediction, and table relation detection would be promising to explore in the future

\begin{acks}
We thank all anonymous reviewers for valuable comments.
This research was supported in part by NSF Grants IIS-1849816.
\end{acks}

\bibliographystyle{ACM-Reference-Format}
\bibliography{references}


\end{document}